\documentclass[printer]{aa} % for a printer version

\pdfoutput=1
\usepackage{graphicx}
\usepackage{subfig}
\usepackage{soul}
\usepackage{txfonts}
\def\fdeg{\hbox{$^\circ$}}
\newcommand{\apo}{APOGEE}
\newcommand{\prr}{Pan-STARRS}

\begin{document}

   \title{Three-dimensional mapping of the local interstellar medium with composite data}

   %\subtitle{I. Overviewing the $\kappa$-mechanism}

 \author{L. Capitanio \inst{1}
          \and
          R. Lallement  \inst{1}
            \and
          J.L. Vergely \inst{2}
           \and
          M. Elyajouri \inst{1}
          \and
          A. Monreal-Ibero \inst{1}
 }

\institute{GEPI, Observatoire de Paris, PSL Research University, CNRS, Universit\'e Paris-Diderot, Sorbonne Paris Cit\'e, Place Jules Janssen,
92195 Meudon, France\\
              \email{letizia.capitanio@obspm.fr}
             \and
             ACRI-ST, Route du Pin Montard, Sofia-Antipolis, France
}

\date{Received xx, 2017; accepted xx, 2017}

% \abstract{}{}{}{}{} 
% 5 {} token are mandatory
 
 \abstract
{Three-dimensional maps of the Galactic interstellar medium are general astrophysical tools. Reddening maps may be based on the inversion of color excess measurements for individual target stars or on statistical methods using stellar surveys. Three-dimensional maps based on diffuse interstellar bands (DIBs) have also been produced. All methods benefit from the advent of massive surveys and may benefit from Gaia data.}
{All of the various methods and databases have their own advantages and limitations. Here we present a first attempt to combine different datasets and methods to improve the local maps.}
%With more detailed structural and chemical information, we learn how our Milky Way evolved, but we could also use directly the extinction
%for several data analysis. 
{We first updated our previous local dust maps based on a regularized Bayesian inversion of individual color excess data by replacing Hipparcos or photometric distances with Gaia  {Data Release 1} values when available. Secondly, we complemented this database with a series of $\simeq$ 5,000 color excess values estimated from the strength of the $\lambda$15273 DIB toward stars possessing a Gaia parallax. The DIB strengths were extracted from SDSS/APOGEE spectra. Third, we computed a low-resolution map based on a grid of Pan-STARRS reddening measurements by means of a new hierarchical technique and used this map as the prior distribution during the inversion of the two other datasets.}
%To design them there are different techniques; we invert a color excess map, using single linesights and distance 
%coming from 
%first data release Gaia. 
{The use of Gaia parallaxes introduces significant changes in some areas and globally increases the compactness of the structures. 
Additional DIB-based data make it possible to assign distances to clouds located behind closer opaque structures and do not introduce contradictory information for the close structures. A more realistic prior distribution instead of a plane-parallel homogeneous distribution helps better define the structures. We validated the results through comparisons with other maps and with soft X-ray data.}
{Our study demonstrates that the combination of various tracers is a potential tool for more accurate maps. An online tool makes it possible to retrieve maps 
and reddening estimations (http://stilism.obspm.fr).}
%To do this inversion we start from a map designed with the excess of colour from Panoramic Survey Telescope and Rapid Response System (\prr) survey. 
%We use ~28000 sightlines with determined color excess, limiting distance at 2500 pc from Sun, from Str$\ddot{o}$mgrem catalogs, 
%Geneva photometric database, Geneva-Copenhagen survey, \apo{} survey and some open cluster. The color excess is derived for the 
%\apo spectra using the DIB 1.5273 $\mathrm{\mu m}$ equivalent width as proportional index.
%The inversion technique is similar to previous maps, using regularized Bayesian approach. 
%The results are available on the web site www.stilism.gepi.obspm.fr}
%\end{abstract}
%  \end{@twocolumnfalse}
%]

   \maketitle
%
%-------------------------------------------------------------------

\section{Introduction}

Three-dimensional (3D) maps of the nearby and distant Milky Way interstellar medium (ISM) are useful multipurpose tools have been only recently developed at variance with the extremely detailed, 
multiwavelength 2D emission maps or spectral cubes that have been accumulated over decades. As  a matter of fact, a revolution is underway that dramatically 
changes the situation: photometric, spectrophotometric, and spectroscopic massive stellar surveys have started to  provide extinction and absorption 
data that, coupled with parallax or photometric star distances, bring the third dimension for the ISM, i.e., the distance to the clouds. Thanks to 3D maps of the interstellar matter, structures overlapping on the sky are now converted into distinct clouds distributed at different distances, which will progressively allow emission components to be assigned a well-located production site. 

Three-dimensional maps or pseudo-3D maps have already been produced based on various kinds of stellar data; that is mainly photometric extinction but also diffuse interstellar bands (DIBs) and  {spectral} lines of gaseous
 species  \citep{Marshall06, Vergely10, vanloon13, Welsh10, lallement14, Schlafly14, Sale14, kos14, Schultheis14, zasowski15, Green15,Schlafly15,Rezaei16}. Various 
 methods have been used to synthesize the distance-limited data, including full 3D tomographic inversion of individual sightline extinctions \citep{Vergely01,SaleMag14}, 
 statistical methods based on photometry, and stellar population synthesis or color-color diagrams; see the \cite{EWASS15} proceedings for recent developments in this field. In the 
 coming years, maps of increasing quality are expected thanks to continuing or new stellar surveys, on the one hand, and precise Gaia parallaxes, on the other hand. 

Each mapping technique has its advantages and limitations. Statistical methods require large amounts of targets and are appropriate at large distances, while the synthesis of individual 
line-of-sight data may work closer to the Sun. Near-infrared (NIR) or infrared (IR) surveys allow one to go deeper in cloud opacities compared to optical data but are more difficult to correct for telluric 
emissions and absorptions. Maps based on DIBs may slightly suffer from  the fact that DIB carriers tend to disappear in dense cores of clouds \citep{Lan15}, thereby potentially reducing 
the resolution. Nevertheless, there are numerous DIBs and their combinations may become powerful tools to probe the physical state of the encountered media. Whatever the technique, a major difficulty is associated with the decrease with distance of the achievable spatial resolution in the radial direction, due to increasing uncertainties on target distances. Full 3D inversions that combine the radial and ortho-radial information may partly overcome this difficulty  provided they are based on massive datasets, however at the price of strongly increased computing time (see, e.g., \cite{SaleMag14}).

We present a first attempt to address these limitations by means of a combination of three different datasets as follows: (i) individual color excess measurements from broad- and narrowband photometry of nearby stars (mostly $\leq
$500pc), (ii) NIR DIB equivalent widths  {(EWs)} measured in individual spectra of more distant stars (500-1500pc), and (iii) color excess measurements at larger distance (up to 3kpc) 
based on a 
statistical analysis of multiband photometric data. Our purpose is to test such an association and its potential for improved 3D mapping of nearby or distant  ISM. 
 The first two datasets are 
merged by means of a conversion from DIB strengths to E(B-V) color excesses. A Bayesian inversion is applied to the resulting composite database. 
The large distance color excess data are 
used to build a very low-resolution 3D distribution that is subsequently used as the prior distribution in the Bayesian inversion to replace a plane-parallel homogeneous distribution. 
%This 3D prior is obtained by means of a newly developed hierarchical method allowing to reach a distance-dependent resolution. 

Prior to the combination of these different sources, we analyzed the changes that are induced in the inverted 3D maps based on the first color excess dataset alone when Hipparcos or photometric distances are replaced with new Gaia  {Data Release 1 (DR1)}  parallax distances  {(the TGAS catalog)}, when available.

In section 2 we describe the three datasets used in the inversions. In section 3 we briefly recall the Bayesian technique developed in \cite{Vergely10} and \cite{lallement14} (hereafter LVV) and the changes introduced in the method. We also describe the hierarchical method developed to build the large-scale prior distribution. Section 4 describes the 3D map evolution introduced successively using Gaia distances, the addition of the DIB-based data, and finally the large-scale prior map. Section 5 discusses uncertainties and shows comparisons between the resulting final map and other measurements of cloud distances or cavities. Section 6 discusses the results and perspectives of such composite maps.

 %To study how neighbourn in the Milky Way, its quite important have a general idea about the interstellar medium structure. We know that the Sun
%is in Local Bubble, we are in a small branch in our galaxy and we can recognize some supernova events with little boubble in the local interstellar medium.
%To map this medium we cannot directly measure the distances, as the stellar maps, so we need some kind of inversion techniques. There are three most important: statistical, simulation-based, for single line of sight. The first one use large amount of data and have in GREEN **** the best example 
%at this moment. Second one is fed from a sintetical model of stellar distribution in our galaxy, ***. 
%We use the less one, to get a hight detailed local map. 
%We call plane the surface with at longitude 0, not the real galactic plane.
%In thise paper we prensent the data used in the inversion in Section \ref{data}, then the inversion technique details in Section \ref{inversion} and 
%the end we show the most important improvement got with this map and its limits

%To read this paper is quite important read also \cite{rosine:oldmap}
\section{Data}
\label{data}

\subsection{Individual color excess data}\label{dataold}

\begin{figure*}
\centering
\includegraphics[width=0.8\textwidth]{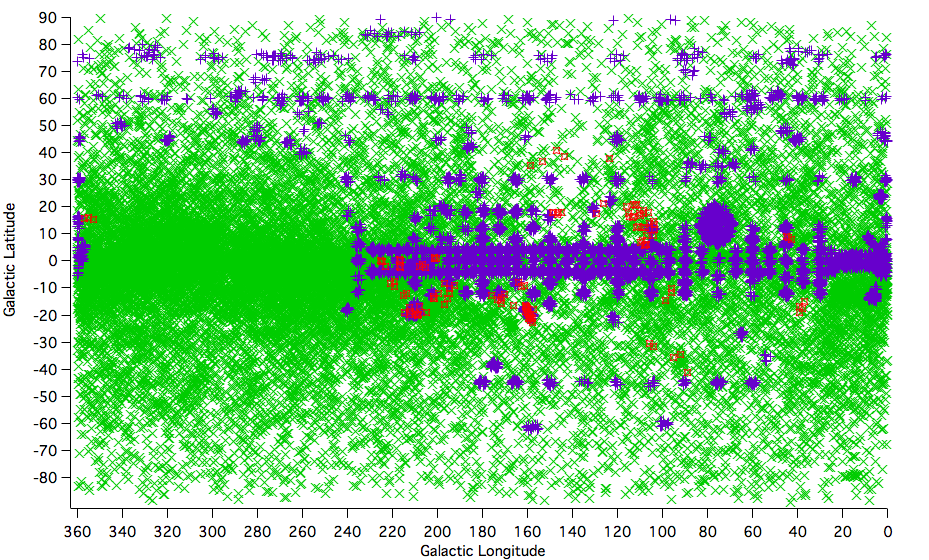}
\caption{Location of the targets from the LVV catalog compilation (green crosses) and of the \apo{} target stars retained for the DIB extraction and the inversion (violet signs). Coordinates are Galactic.  Also shown are the directions of the molecular clouds of the \cite{Schlafly14} catalog mentioned in section \ref{molcloudcomp} (red signs).} 
\label{Targets1and2}
\end{figure*}

 We started with the reddening dataset of $\simeq$22,500 sightlines compiled and inverted by LVV. It comprises four different catalogs  {that were} cross-matched and homogenized;  for more details about the stellar types and characteristics of the calibration method for each separate dataset, see \cite{Vergely10, Cramer99, Nordstrom04, Dias12}. 
 Most target stars are nearby (d$\leq$500pc), cluster data only correspond to larger distances. The four catalogs are complementary in terms of sky coverage. Of the target distances used for the inversion presented in LVV, 
 75\% were photometric. For one-quarter of these distances the Hipparcos distance  {was} preferred.  
 In the four inversions presented here we kept the same error bars on the reddening as those used by LVV.  
 The only difference in the LVV series of targets is the removal of $\simeq$400 targets with photometric distances that had no catalog identifications. 
 The coordinates, brightnesses, estimated distances, and reddening reasonably of these  $\simeq$400
targets correspond to those of nearby stars. 
Therefore, these targets {were not excluded} for the inversion presented in LVV. 
 However, as we discuss in section \ref{inversions}, after the addition of new, more distant APOGEE targets and the subsequent mapping improvement we realized that for these particular targets the color excesses were significantly above other data in the same area. For this reason, we conservatively excluded the targets. Fig. \ref{Targets1and2} shows the distribution of the remaining targets over the sky.

As shown in LVV, the distribution of the targets is highly biased toward low reddening regions in the sky. This had strong consequences on the resulting maps; cavities surrounding the Sun were well defined along with the boundaries of the first encountered dense clouds. On the other hand, clouds located behind foreground 
 dense structures were either not mapped or were very poorly mapped. 
  
\subsection{Measurements of DIBs and DIB-based color excess estimates}\label{dibdata}

Except for the recent identification of ionized fullerenes C$_{60}^{+}$, the carriers of the DIBs are still unknown, and it may seem strange to use absorptions due to unknown species for mapping purposes. 
However the link between DIB strengths and the columns of interstellar matter is strong and successful DIB-based 3D mapping has now been carried out (see Sect. 1). 
Up to now  such mapping used solely DIBs. At variance with these works, here we attempted to merge color excess data and DIB absorption data and perform a unique inversion of a composite dataset. 
This requires a conversion of DIB equivalent widths into reddening (or the inverse); because none of the DIBs have been shown to be perfectly correlated with the reddening, such a mixing of color excesses computed in so many different ways--- for example, with photometric and spectrometric techniques---is potentially hazardous.  
On the other hand, initial comparisons between DIB equivalent widths and reddening radial profiles in the same sightlines have demonstrated that their similarity \citep{Puspitarini15} and global maps also have strong similarities \citep{kos14}. 
Additionally, it is possible to account for the variable E(B-V)/DIB ratio by increasing error bars on the estimated E(B-V), which we have carried out in this work.

\subsubsection{Data}
We used the results of the SDSS-III Apache Point Observatory Galactic Evolution Experiment (APOGEE), which is one of the four Sloan Digital Sky Survey III \citep[SDSS-III;][]{Eisenstein11,Aihara11} experiments. The APOGEE uses a 300-fiber multiobject spectrograph working in the near-infrared ($H$-band; 1.51- 1.70~$\mu$m) at high spectral resolution \citep[R$\sim$22\,500;][]{Wilson10}. 
Specifically, we downloaded spectra from the SDSS data release\footnote{http://www.sdss.org/dr12/} \citep[DR12;][]{Alam15}, which provides all the data taken between April 2011 and July 2014.\\ 
In addition to the $\sim$160\,000 scientific targets selected from the source catalog of the Two Micron All Sky Survey \citep[2MASS; ][]{Skrutskie06} and distributed across all Galactic environments, APOGEE contains a sample of $\sim$17\,000 spectra of hot stars, called telluric standard stars (TSSs). The TSSs form the basis for the correction of the telluric absorption lines across all field types (see \cite{Zasowski13}, for a detailed description of the different target classes). These stars are the bluest stars on a given APOGEE plate with a magnitude in the range $5.5\le$ H $\le11$ mag and are therefore hot and bright stars with spectra that are most often featureless.\\ 
The APOGEE products contain the decontaminated spectra and synthetic stellar spectra that provide the main stellar line locations and relative depths and widths. Both  {were} used by \cite{Elyajouri16} to extract a catalog of $ \lambda$15273 DIB measurements from 6700 TSS spectra. Further details on the sample can be found in \citet[][ and references therein]{Elyajouri16}. Weaker DIBs have been further studied based on the same TSS spectra \citep{Elyajouri17}. Here we focus on the strongest 15273 \AA{} DIB with the aim of extracting as many accurate measurements as possible from both scientific targets and TSSs possessing a Gaia parallax.\\

We analyzed  the series of 25\, 409 DR12 spectra of targets belonging to the Tycho catalog to determine the presence or absence of a detectable DIB at 15273 \AA\  and measure its equivalent width (resp. an upper limit) in case of positive (resp. negative) detection. We took as a starting point the methodology presented by \cite{Elyajouri16} for TSSs spectra. We kept the same fully automated fitting technique and method of classification but we adapted the fitting constraints to take into account the inclusion of targets cooler than the TSSs. The code was developed using the IGOR PRO environment 2.\\

\subsubsection{Fitting method}
We restricted the fit to the spectral range $ [15\,260-15\,280]\AA$ in the vicinity of the DIB. This range is wide enough to ensure an adequate sampling of the neighboring Bracket 19-4 stellar line at 15264.7 \AA. We used for the fit the Levenberg-Marquardt algorithm implemented in IGOR PRO\footnote{\texttt{http://www.wavemetrics.com}}.\\

We fit the data to a combination of an adjusted synthetic spectrum, DIB model, and smooth continuum as shown in equation \ref{equfit}. Below we provide information specific to each of the equation variables as follows: 
\begin{equation}
M_\lambda= [S_\lambda]^\alpha \times DIB[\sigma,\lambda,D]  \times 
(1+[A] \times \lambda)
\label{equfit}
,\end{equation}

$\bullet$ $[S_\lambda]^\alpha$, an adjusted stellar spectrum, where S$_\lambda$ is the stellar model provided by APOGEE. We included the scaling factor $\alpha$ to adjust the model stellar line depths to the data.

$\bullet$ $DIB[\sigma,\lambda_c,D]$, the DIB shape was modeled as a Gaussian function allowed to vary in strength, depth, and velocity shift with three free parameters associated with its Gaussian RMS width ($\sigma$), central wavelength ($\lambda_c$), and depth ($D$).

$\bullet$ $(1+[A] \times \lambda)$, a 1-degree polynomial function introduced to reproduce the continuum around the DIB as close possible.\\

An example illustrating our fitting procedure is shown in Figure \ref{apofit}.

\subsubsection{Selection criteria and error estimates}

Residuals, {computed as differences between data and model}, were used to obtain three measurements of the standard deviations RA, RB, and RC. To do so, we performed the local fit and a second fit over the whole spectral range covered by APOGEE using the same fit function as that presented in equation \ref{equfit}.\\

%The DIB is fitted to ensure an optimum placement of the continuum in the DIB region and subsequently everywhere. As a matter of fact, omitting the DIB could bias the continuum in its spectral region and react on the whole fit.

We selected the standard deviations RA, RB, and RC of the fit residuals, as follows: 

$\bullet$ RA from the local fit was calculated in a region A =  [$\lambda_{c}$ - 10 $\AA\ $, $\lambda_{c}$+ 10 $\AA\ $]  close to the DIB. 

$\bullet$ RB was calculated over a spectral range B= [15890-15960] \AA,\ which is free of DIB absorption,  strong stellar lines, and telluric residuals.

$\bullet$ RC was obtained from region C= [15200-15250] \AA\ relatively close to the DIB, and is also free of DIB and potentially contaminated by stellar residuals.\\ 

These are used for the selection criteria shown in Table \ref{tablecont} and also used below to estimate the uncertainty on the DIB equivalent width. The uncertainty is conservatively estimated based on the following formula:

\begin{equation}
\label{err}
\delta_{EW} =  \sqrt{ 2\pi} \:  \sigma\: max(RA,RB,RC)
,\end{equation}

where $\sigma$ is the Gaussian RMS width that results from the fit. The quantity max (RA,RB,RC) is the maximum of the three standard deviations. \\

\begin{figure*}
\centering
\subfloat[][]
{\includegraphics[width=.33\textwidth]{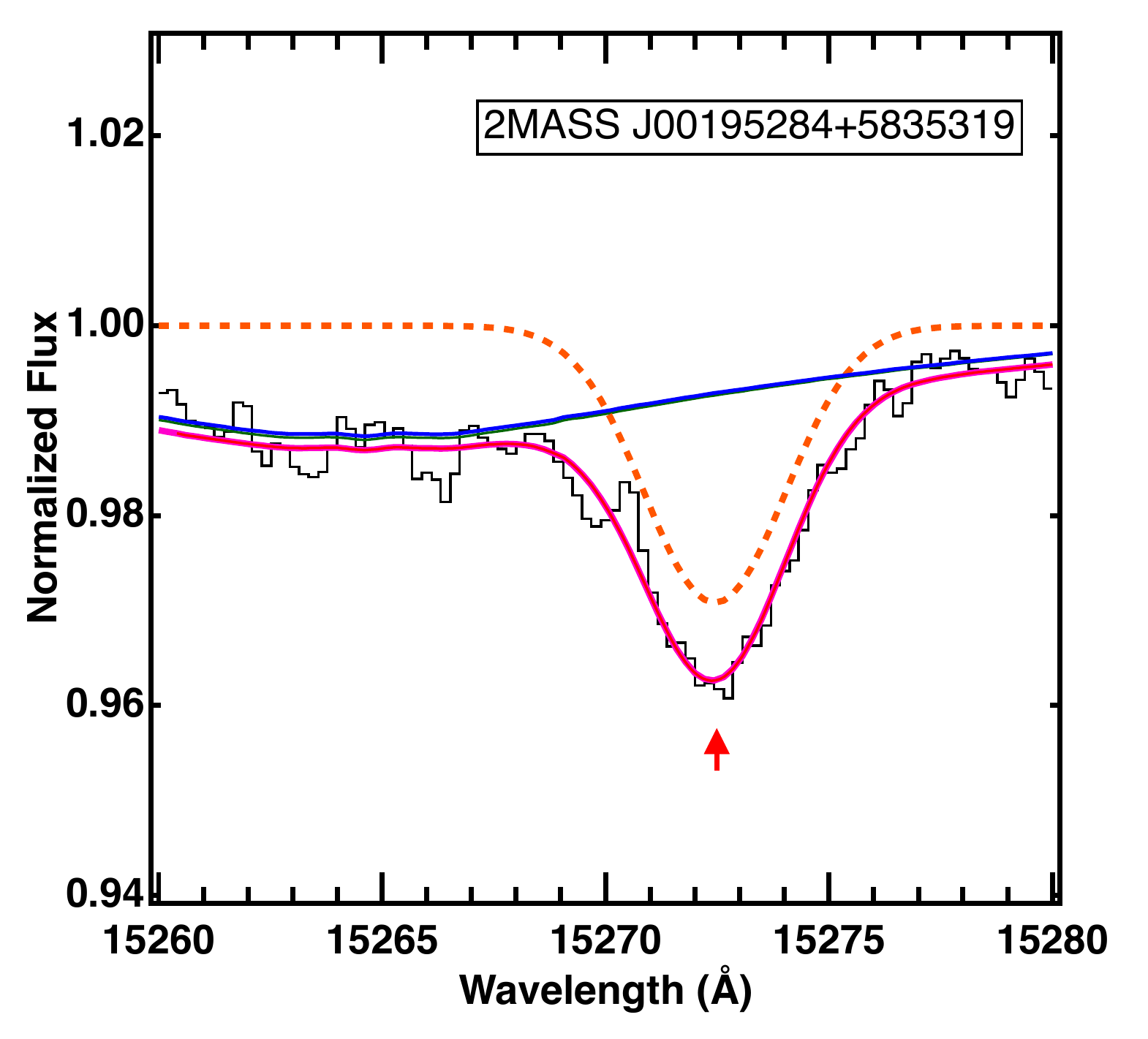}} 
\subfloat[][]
{\includegraphics[width=.33\textwidth]{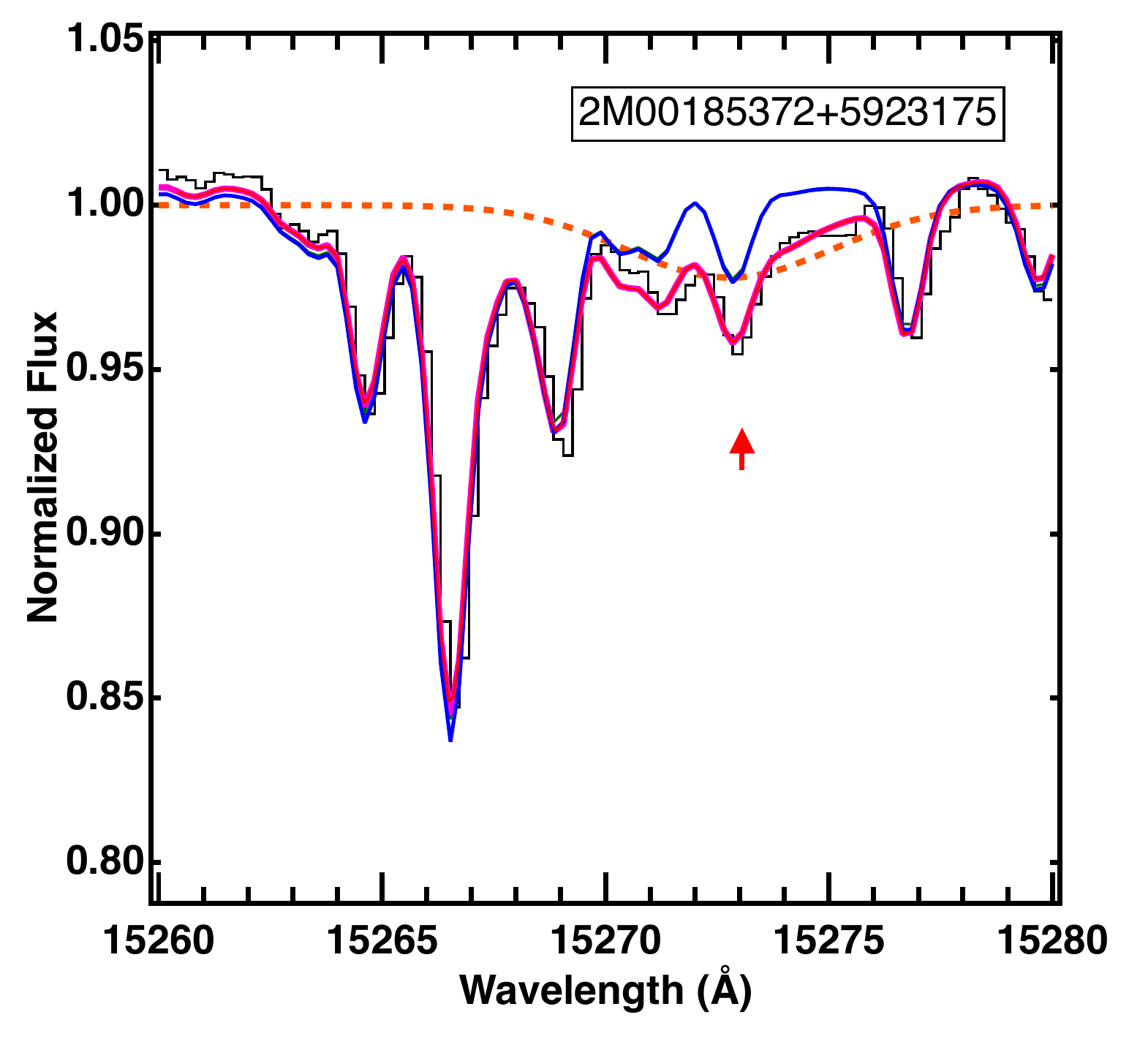}} \\
\subfloat[][]
{\includegraphics[width=.33\textwidth]{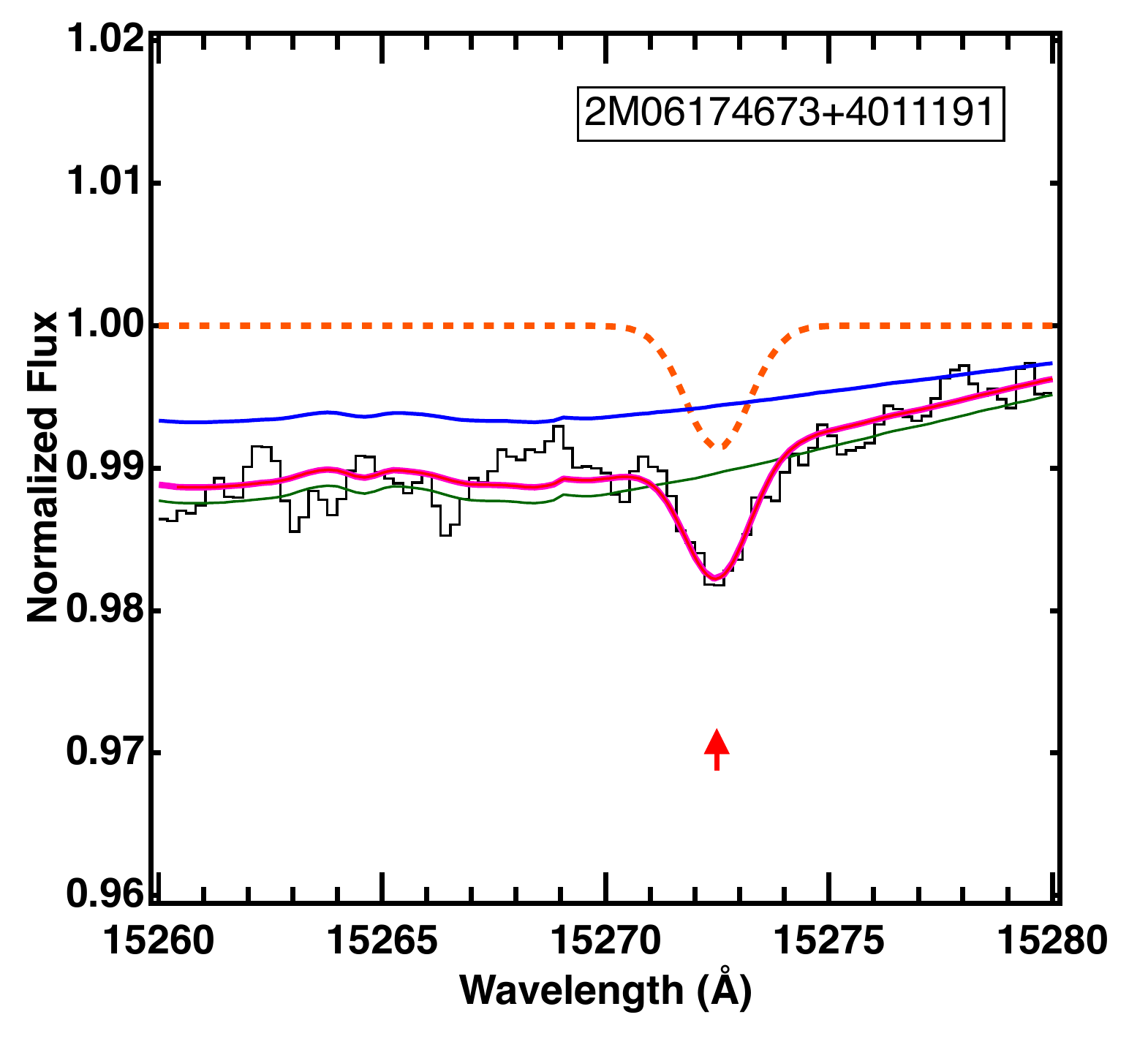}} 
\subfloat[][]
{\includegraphics[width=.33\textwidth]{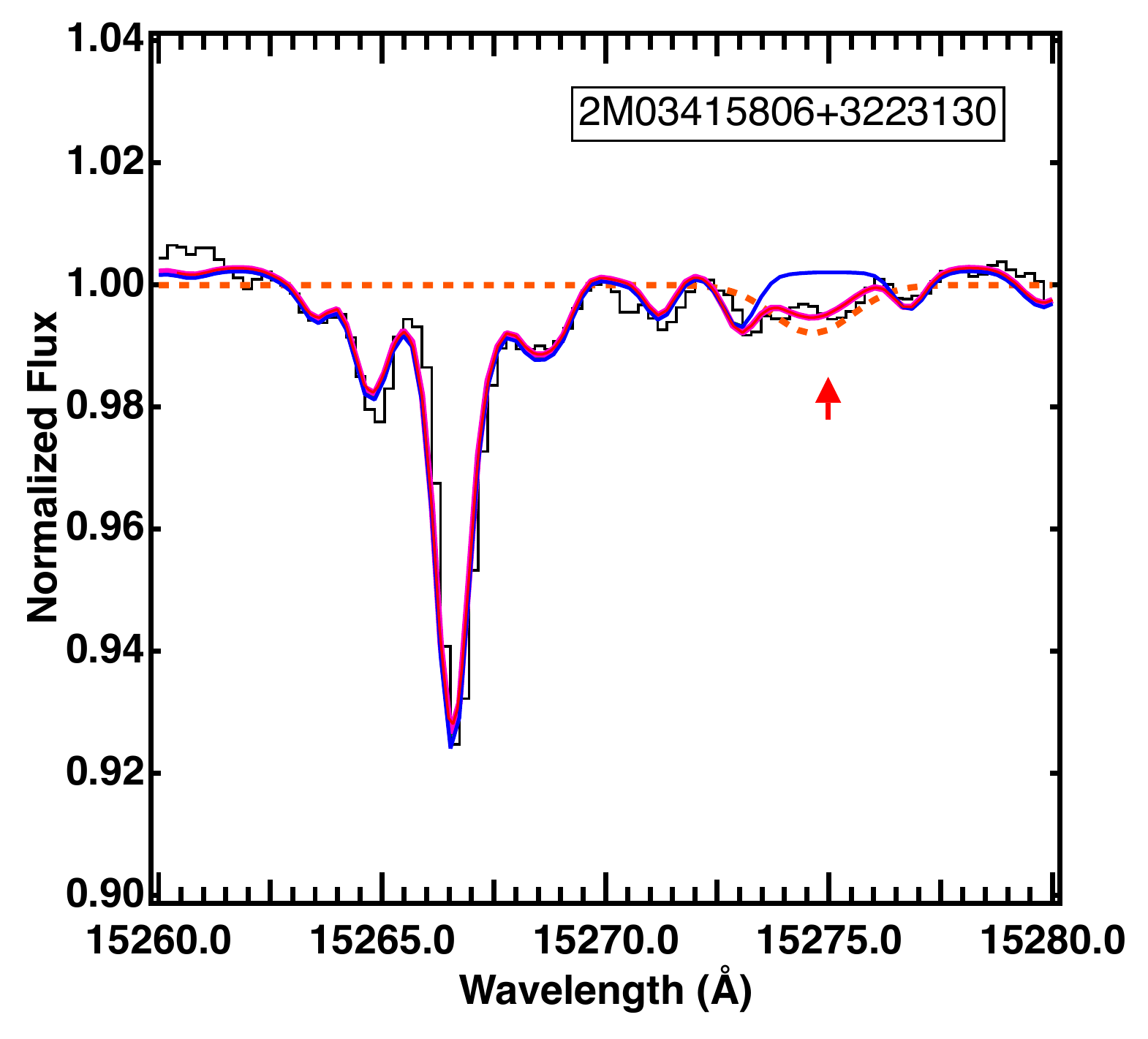}} \\
\subfloat[][]
{\includegraphics[width=.33\textwidth]{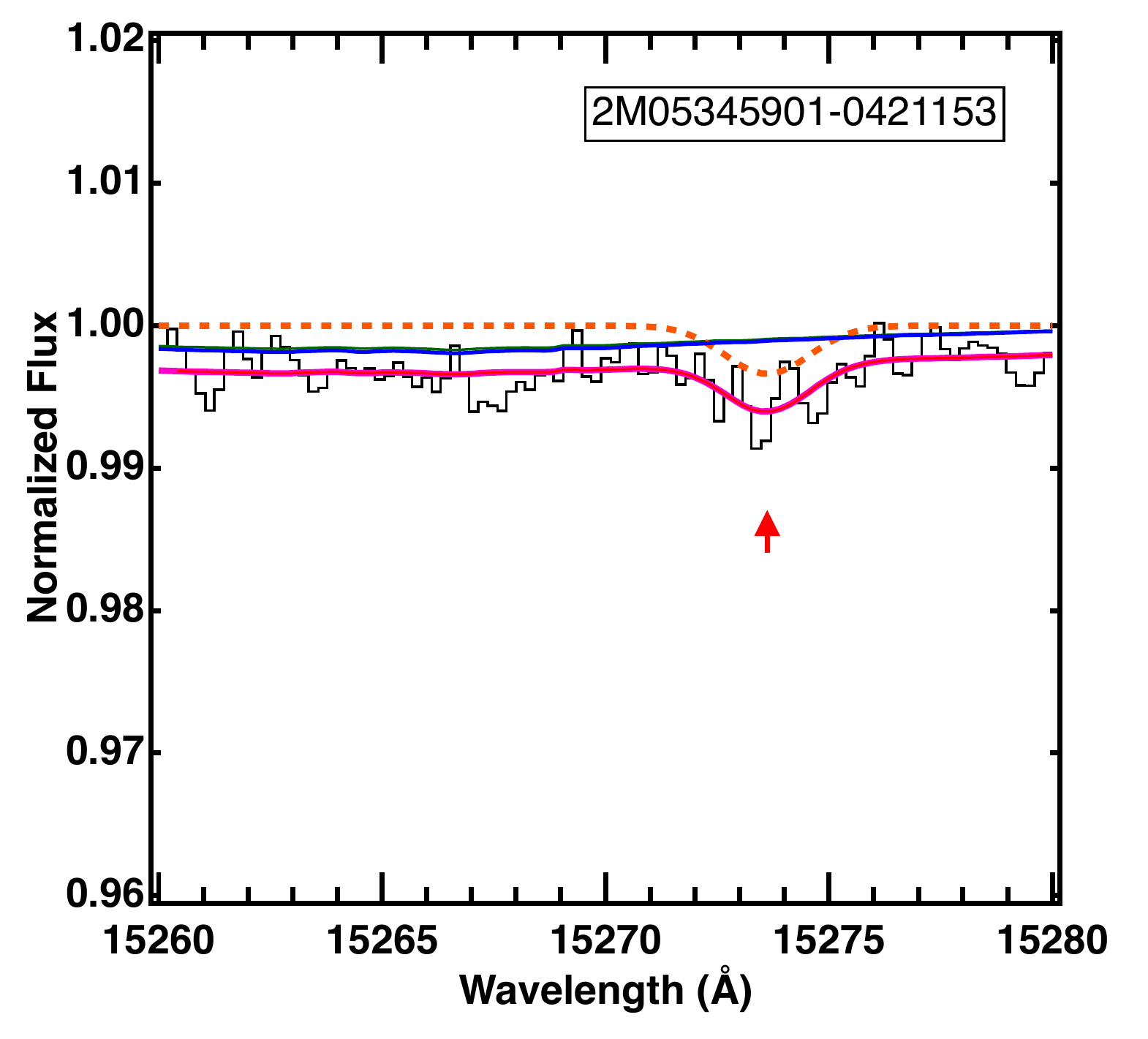}}
 \caption{Illustration of the various categories of DIB extraction: a) Detected; b) Recovered; c) Narrow; d) Recovered and Narrow; e) Upper limit. APOGEE spectra are shown with solid black curves. The initial stellar model provided by APOGEE and the model obtained after application of the scaling factor are shown in solid blue and green respectively (note that in many cases they are very similar). The DIB absorption alone is represented in orange. The solid magenta curves represent the final stellar+DIB modeled spectrum.}
\label{apofit}
\end{figure*}

Table \ref{tablecont} shows the fitting constraints that were used to select our final sample. They are somewhat different from the criteria of \cite{Elyajouri16}. In particular, we no longer used HI 21 cm data to limit the width of the DIB, but we used limits based on the FWHM histograms built in this previous work.  Representative examples illustrating our fitting procedure and the various categories are shown in Fig. \ref{apofit}.\\

\begin{table*}
\centering                          % used for centering table
\begin{tabular}{l l l l}  
\hline\hline 
Flags& $\lambda_{c}$&Width&Depth\\
& $\AA$ & $\AA$ & $\AA$ \\ 
\hline\hline
Detected (4861)&$ [15\,268-15\,280]$&$[1.4 - 5.5]$&$[1.5\times$ max(RA,RB,RC) $- 0.1]$\\
%8 : dib detected
%15268<lambda_centrale<15280 &  1.4=<width< 5.5 &  1.5*sdev3?depth< 0.1 
% choisir Figs : detected1, detected2, detected3
Narrow DIBs (364) &$ [15\,260-15\,280]$ &$[1 - 1.4]$&$[1.5\times $ max(RA,RB,RC) $- 0.1]$\\
%4 : small dib (Ètroites) 
%15260<lambda_centrale<15280 &  1<width< 1.4 &  1.5*sdev3?depth< 0.1 
% choisir Figs : narrow1, narrow2
Recovered DIBs (2515) &$ [15\,268-15\,280]$ &$[1.4 - 5.5]$&$[2\times$ RA $- 1.5\times$ max(RA,RB,RC)]\\
%7 : dib rattrapee
%15268<lambda_centrale<15280 &  1.4=<width< 5.5 & 2*SDEV_dib ?depth< 1.5*sdev3
% choisir Figs : rattrap1, rattrap2, rattrap3
Narrow recovered DIBs (420) &$ [15\,268-15\,280]$&$[1 - 1.4[$&$]2\times$  RA  $- 1.5\times$ max(RA,RB,RC)]\\
%10 : dib small rattrapee ds l?intervalle le plus probable
%15268<lambda_centrale<15280 &  1=<width< 1.4 &  2*SDEV_dib ?depth< 1.5*sdev3
% choisir Figs : narrow_rec1, narrow_rec2, narrow_rec3
Upper limit (1753) & $ [15\,268-15\,280]$ & $\leq$1.4 & $\leq$ $2\times $RA\\
%12 : no certain, micro dib 
%15268<lambda_centrale<15280 &  width< 1.4 & depth< 2*SDEV_dib
\hline
\end{tabular}
\caption{Fitting constraints of $\lambda$15273 NIR DIB for the selected APOGEE spectra. RA, RB,  RC are the standard deviation in three spectral ranges A =  [$\lambda_{c}$ - 10$\AA$, $\lambda_{c}$+ 10$\AA$], B= [15890-15960] $\AA$ and C= [15200-15250] $\AA,$  respectively.}             
 \label{tablecont}       
\end{table*}

\subsubsection{Final data selection for the inversion}
Equivalent widths and stellar rest-frame wavelengths are determined from the best-fit parameters in each sightline for the 25,000 attempted spectra of targets belonging to the Tycho catalog. We rejected those cases where the fit failed due to low S/N, an inadequate stellar model, or very strong telluric contamination most of the time, and we retained 9913 spectra following the criteria in Table \ref{tablecont}. The corresponding targets were crossmatched with the targets of the Gaia DR1 catalog \citep{Arenou17} to obtain a catalog of 7800 stars with Gaia distances. 
%We started with the  $ 8,000 E(B-V)s estimated by conversion of 15273\AA\ DIB equivalent widths towards TGAs target stars. 
We added to the DIB measurement an  additional 50\% uncertainty to account for the E(B-V)/EW variability associated with the influence of the physical properties of the encountered media. As a matter of fact, from the most neutral to the most ionized media, this ratio may reach a variability factor of three   {(see, e.g., \cite{Elyajouri17})}. After the additional filtering for our criteria on target distances and distance uncertainties (see Sect \ref{inversold}), only 4886 targets were retained for the present inversion. Other  targets will become useful after the next Gaia data releases. The number of retained targets is small with respect to the initial number of APOGEE spectra of Tycho stars (less than 20\%) and targets from the reddening catalog. 
{Despite the limited number of these retained objects, they are precious} since they allow us to probe more reddened regions, in particular to reveal a second or third rank of clouds, something often precluded using the bright targets measurements in the optical. \\
We converted the DIB EW into a color excess using the \cite{zasowski15} formula
\begin{equation}
EW = 0.102 ~ A_{V}^{1.01} \AA\
\end{equation}
and the classical average total-to-selective ratio A$_{V}$/E(B-V)= 3.1 \citep{Savage79}. The relationship established by  \cite{zasowski15} has been confirmed,  by \cite{Elyajouri16}, to be valid locally based on the closer TSSs. 

\subsection{Subsample of \prr{} color excess measurements}\label{panstarrsdatabase}

\cite{Green15} have used \prr{} and 2MASS photometric measurements toward 800 millions targets to build remarkable 3D dust reddening maps with a very high angular resolution
 (5-15 arc-minutes) covering distances from $\simeq$300 pc to 4-5 kpc. The publicly available maps contain a quality flag that allows to exclude results at small or large distances that 
 are uncertain because of limitations on the number of targets for geometrical or sensitivity reasons. Both surveys were conducted from the Northern Hemisphere.  {We used those data} 
to build a low-resolution  3D \textit{prior} 
 distribution to be used in our Bayesian inversion in replacement of the previous analytical solution corresponding to a homogeneous plane-parallel opacity. 

To do  {this,} we downloaded a grid of \prr{} E(B-V) estimates for lines of sight equally spaced every 0.5 degree in Galactic latitude and longitude and for all distances available from the Pan-STARRS online tool, in this way obtaining 3,071,230
pairs of distance and extinction after exclusion of all data points flagged as uncertain.  {As result of this process,} we do not benefit from the very high angular resolution of the maps, however this information would be lost in our 
approach. In many directions the first 200 pc  are flagged because of the limited number of targets available to define the reddening in the very small solid angle chosen for this 
\cite{Green15} study. However this is where the present inversion of individual data may be advantageous. The method applied to these Pan-STARRS  results is described in Section  \ref{hierarchical}.
 
\section{Inversion techniques}

\subsection{Bayesian inversion of individual sightlines}\label{classicalinv}

The Bayesian inversion technique is based on the pioneering work of \cite{Tarantola82} and its first developments and applications to the ISM are described in detail in \cite{Vergely01, Vergely10, Welsh10, 
lallement14}. Briefly, the inversion optimizes an analytical solution for the 3D opacity distribution through the adjustment of all 
integrated opacities along the target sightlines to the observed color excess data.
 Because the solution is strongly under-constrained, it  is regularized based on the assumption that the 3D solution is smooth. Opacities $\psi$ at two points $(x)$, $(x')$ in space follow correlation kernels: 
 here $\psi_1(x,x')=\exp(\frac{-\|x-x'\|^2}{2\xi_1^2})$ and $\psi_2(x,x')=\frac{1}{\cosh(-\|x-x'\|/\xi_2)}$, where the first (resp. second) term represents the compact (resp. diffuse) structures. 
 The correlation distances $\xi_1$ and $\xi_2$,  {which} roughly correspond to the minimum size of the computed structures, are evidently limited by the average distance between the target stars. 
 Here we have chosen $\xi_1$=15 pc and $\xi_2$=30 pc.  As we discuss in the next sections, 
 such kernels are valid in internal regions where the target distribution is the densest. 
 The model variance, which controls the departures from the prior distribution, is $\sigma_1$=0.8 for the first kernel and $\sigma_2$=1.0 for the second (see LVV for its definition). The \cite{Tarantola82} formalism 
 allows one to compute the least-squares solution iteratively. For the present maps the convergence is reached in about six iterations. Uncertainties on the reddening and target distance are treated in a 
 combined way (see \cite{Vergely01}). 
% have limited the distance to the input targets to 2601 pc, and their height above the Plane to 1000 pc.
The distribution is computed in a volume of 4 kpc by 4 kpc along the \textbf{Galactic plane} and 600 pc along the perpendicular direction. 

Compared to the method used in \cite{lallement14}, an additional treatment of outliers was introduced. 
During each iteration except the first two, in case one measured color excess value is found to be incompatible with the kernels and the surrounding data, its error bar is multiplied by 2. 
This method has proven to solve for data points that have underestimated errors and eliminates oscillations.
The Bayesian approach implemented in \cite{Vergely01} and subsequent works allows us to include an \textit{a priori} knowledge of the ISM 3D distribution. 
This prior distribution can in principle be provided by a model that represents the density variations in the Galactic plane by an analytic law or alternatively by a distribution based on external data. The latter case is fully appropriate to our objective of combining different sources of extinction. As a matter of fact, it is possible to produce density maps by successive refinements following integration of data from different sources in a sequential and non-simultaneous manner. The advantage of such an approach is twofold: first, the amount of data to be integrated at each step is reduced, which makes it possible to reduce computation times and, second, it is easier to see the contribution of each set of data.  

In this perspective, a new development was introduced in our inversion code. Now, it is possible to use any 3D prior distribution,  i.e., we can replace the homogeneous plane-parallel prior that corresponded to an analytical representation (we used an exponential decrease from the Galactic plane) with any arbitrary precomputed distribution.
We used this possibility to test a two-step inversion based on two different datasets as described in section \ref{PSINV}. The construction of the prior distribution is described in the next section.

\begin{figure*}
\centering
\includegraphics[width=8cm]{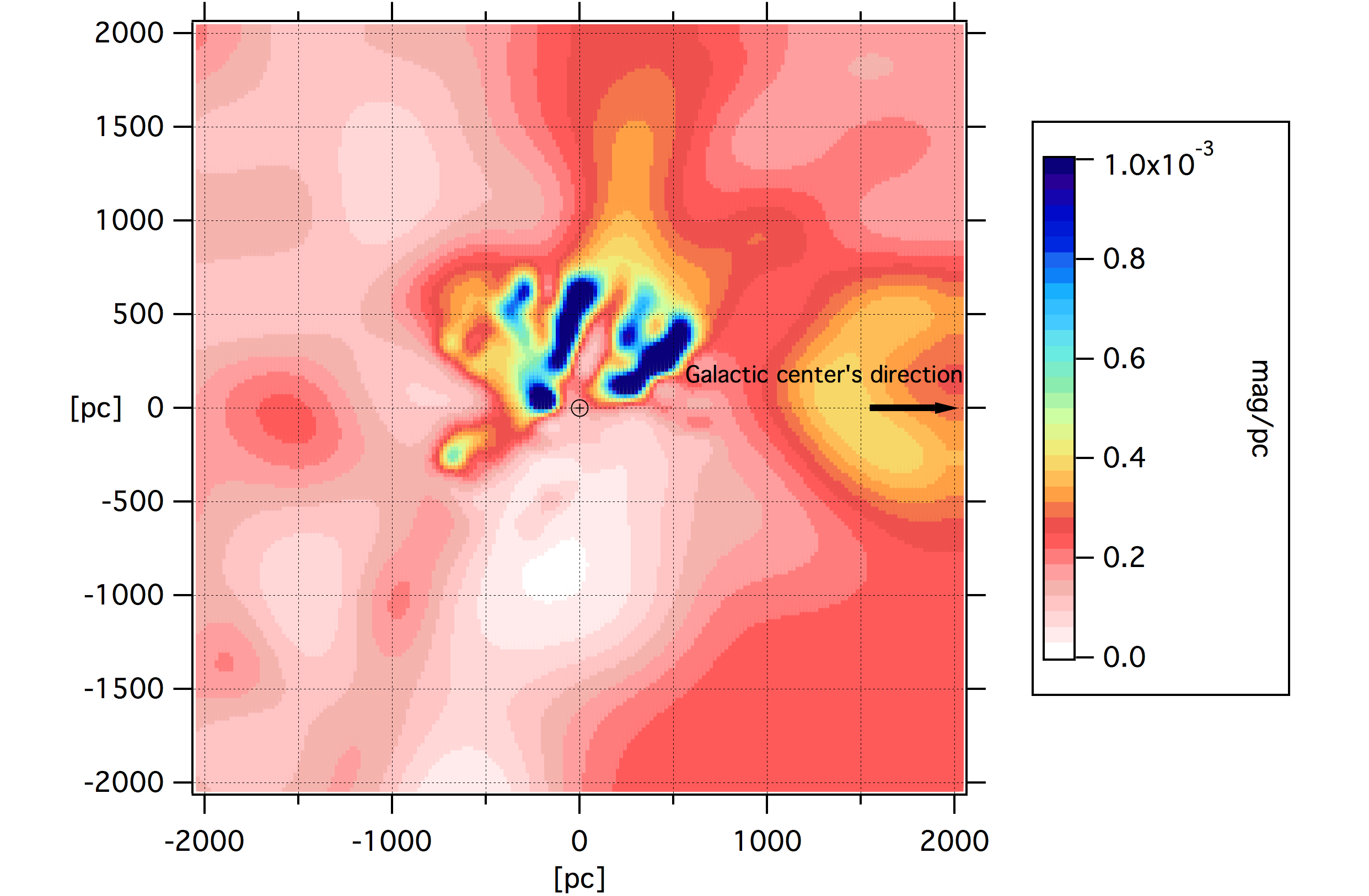}
\caption{Galactic plane cut in the very low-resolution 3D opacity distribution to be used as a prior in the inversion. It is based on Pan-STARRS reddening measurements of \cite{Green15}. 
The Sun is at (0,0) and the Galactic center to the right. Longitudes increase counter-clockwise. Units are mag pc$^{-1}$.}
\label{priorps}
\end{figure*}

\subsection{Large-scale prior distribution based on Pan-STARRS}\label{hierarchical}

We used the \cite{Green15} color excess subsample described in \ref{panstarrsdatabase} to develop a low-resolution opacity model extended to large distances, 
to be used as a prior for the inversion of individual lines of sight as described above. Because this reddening model is based on photometric measurements of hundreds of millions of objects, the \cite{Green15} model 
has a very high angular resolution, whereas the radial resolution is limited by uncertainties on photometric distances. In principle our prior distribution is losing all the details contained in the angular variations,
 first because we start with a coarse angular sampling of 0.5 deg, and also because our full 3D inversion has a spatial resolution limited by the radial distance uncertainties. 

Ideal inversions are based on a homogeneous spatial sampling of targets, i.e., a homogeneous volume density of targets and similar uncertainties on their locations at all distances, 
which allows us to retrieve structures of the same minimal spatial scale everywhere in space. 
But for most surveys, including Pan-STARRS, the achievable resolution is dependent on the distance to the Sun, especially owing to distance-dependent uncertainties on the star locations. 
The sampling of the data in this case does not allow the Nyquist-Shannon criterion to be fulfilled for the same minimal autocorrelation length everywhere;
also, some of the energy attributable to high spatial frequencies is likely to be redistributed in low spatial frequencies, especially at large distances or in the radial directions when the sampling is weak. 
Therefore, we need an approach that allows us to retrieve only the large scales at large distances and all large and small scales at short distances. 
We  implemented a hierarchical approach that allows us to achieve this goal in an iterative way by progressively updating the opacity model. 
Schematically, we first constructed a low spatial frequency map in a sphere with a large radius (here 5 kpc). 
In a subsequent step, we estimated spatial frequencies that are higher by a factor of 2, however this time in a smaller sphere (for instance 3 kpc) by removing 
data at large distances for which the available information does not make it possible to reach these frequencies using the Nyquist-Shannon criterion. 
The information at large distances is kept under the form of the prior, which is simply the distribution derived from the previous step. 
This procedure is repeated for increasingly high frequencies until their maxima compatible with the radial sampling (reached at small distances) are obtained.  In practice we proceeded as follows:
\begin{enumerate}
\item The \prr{} reddening data subsample (see section \ref{panstarrsdatabase}) is converted into local opacities $\rho_0$ through derivation w.r.t. the radial distance R,\\
 $\rho_0= {\delta_{E(B-V)} }/ {\delta_R}$.  
\item A spatial scale S of restitution of the opacity is fixed, starting with the (large) scale that can possibly be restored on the whole map. The average opacity is computed in a 3D grid of  
boxes of size SxSxS. The resulting average opacity is thus dependent on the scale.
\item Based on these average opacities, maps of smoothed opacities are computed by Bayesian interpolation with a smoothing kernel twice as large as S. 
\item We then move on to a scale twice smaller. S/2 x S/2 x S/2 boxes are built and the average opacity is calculated for each box. 
  The dynamics of the average opacity for this set of boxes is stronger. We suppress the boxes that do not fulfill the Nyquist-Shannon criterion for S/2.
\item A Bayesian interpolation is carried out again with a smaller smoothing kernel and using the previous map as prior.
\item Again, the scale is divided by two and steps 2) and 3) are performed again  for this new scale, etc.
\end{enumerate}
A planar cut along the \textbf{Galactic plane} in the resulting 3D prior is shown in Fig. \ref{priorps}. The Sun is at center and Galactic longitude increases counter-clockwise.
As expected, far from the Sun we can only restore the very large scales. 
On the contrary, one can reach scales on the order of 100 pc within the first few hundreds parsecs. Because Pan-STARRS data are recorded from the Northern Hemisphere, 
the prior in the fourth (bottom-right) quadrant appears homogeneous at large distances from the Sun.
This map can be used as a prior in the inversion of the individual lines of sight for which accurate distances are available, 
with the goal of improving estimates of opacity fluctuations and refining the distance assignment of the structures (see Sect \ref{PSINV}).

%--------------------------------------------------------------------
\begin{figure*}
\centering
\subfloat[][A]
{\includegraphics[width=6.0cm]{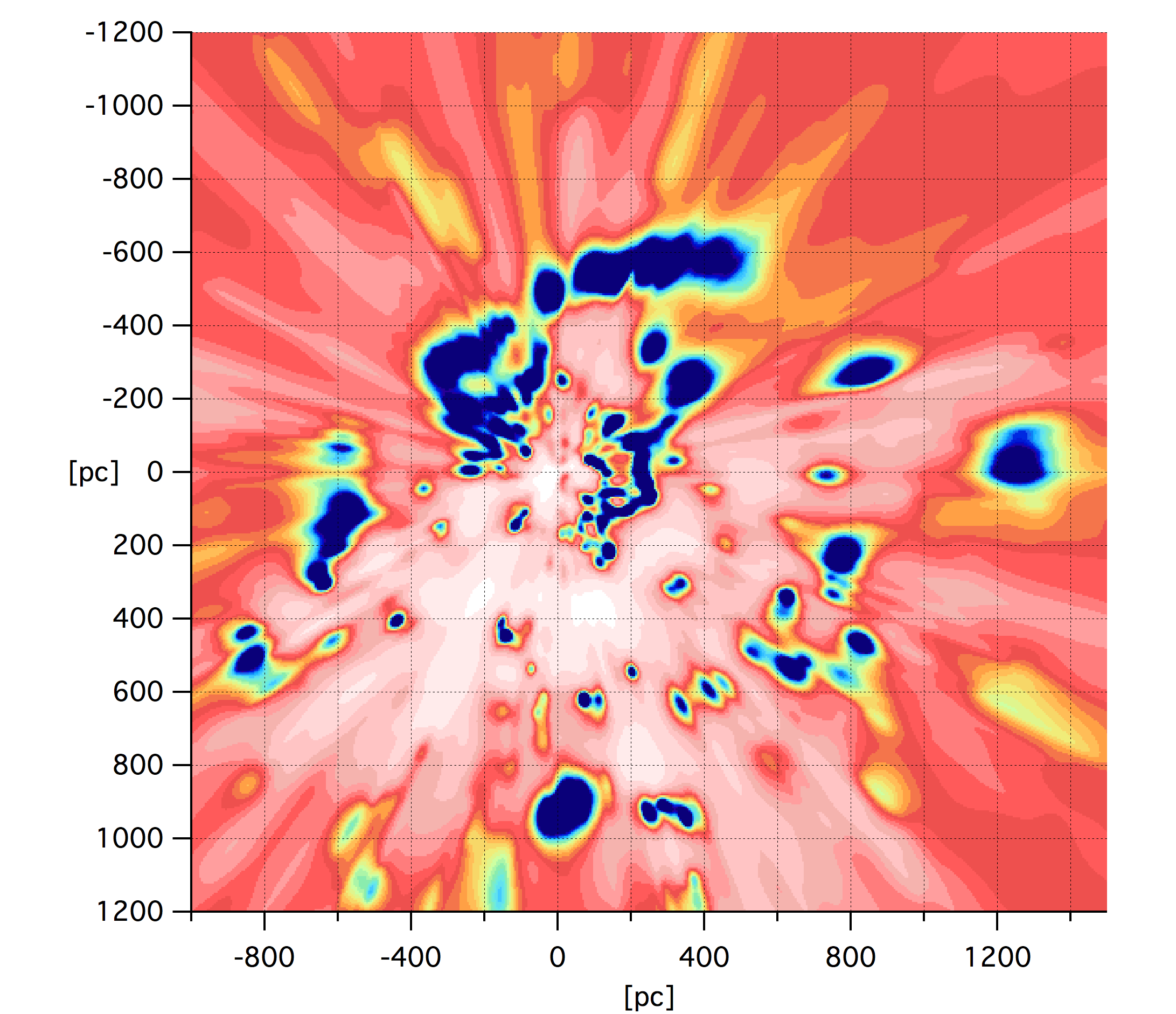}}
\subfloat[][B]
{\includegraphics[width=6.0cm]{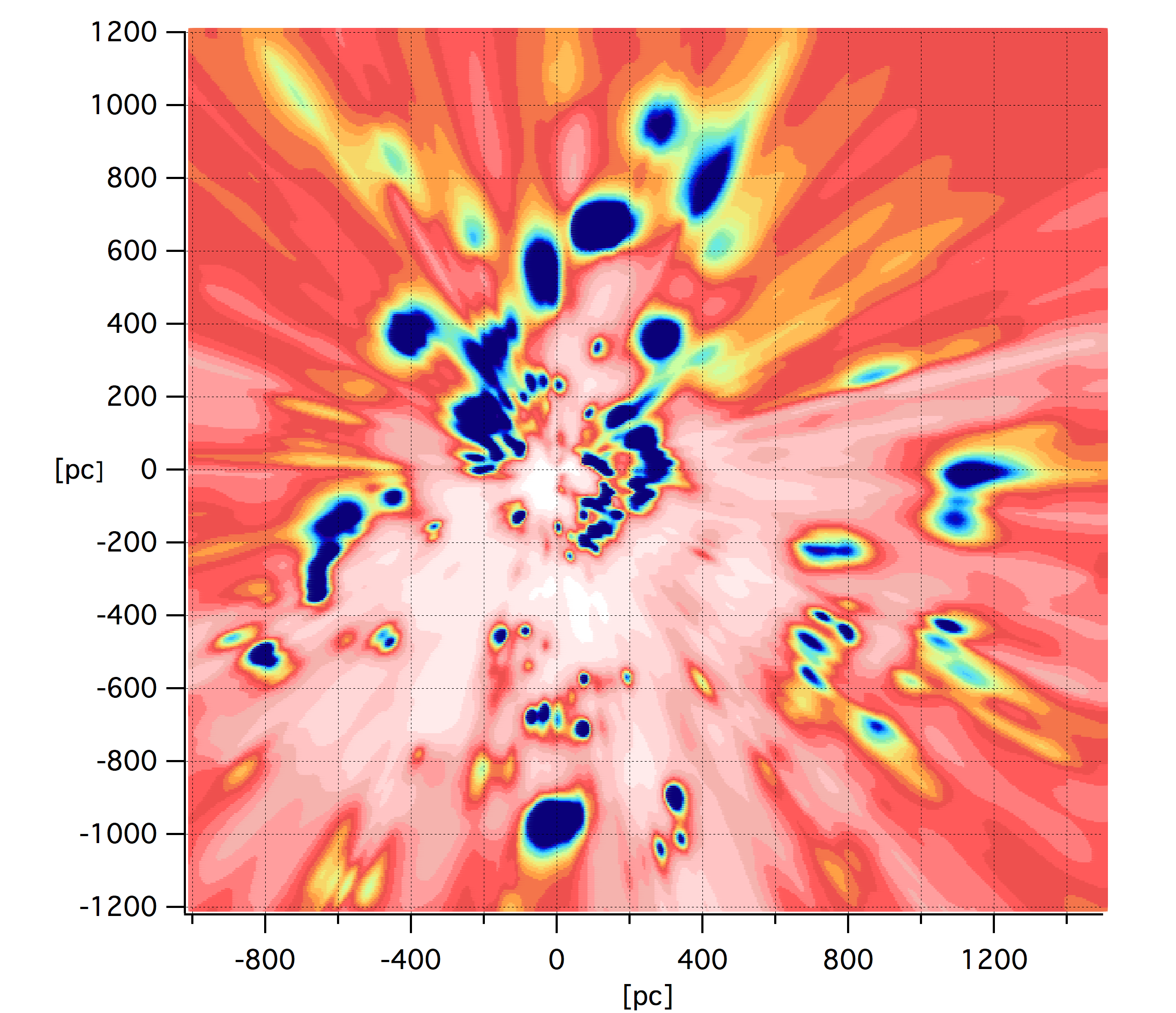}}\\
\subfloat[][C]
{\includegraphics[width=6.0cm]{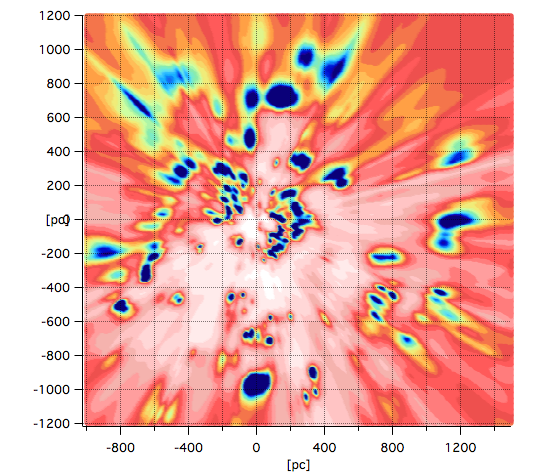}}
\subfloat[][D]
{\includegraphics[width=6.0cm]{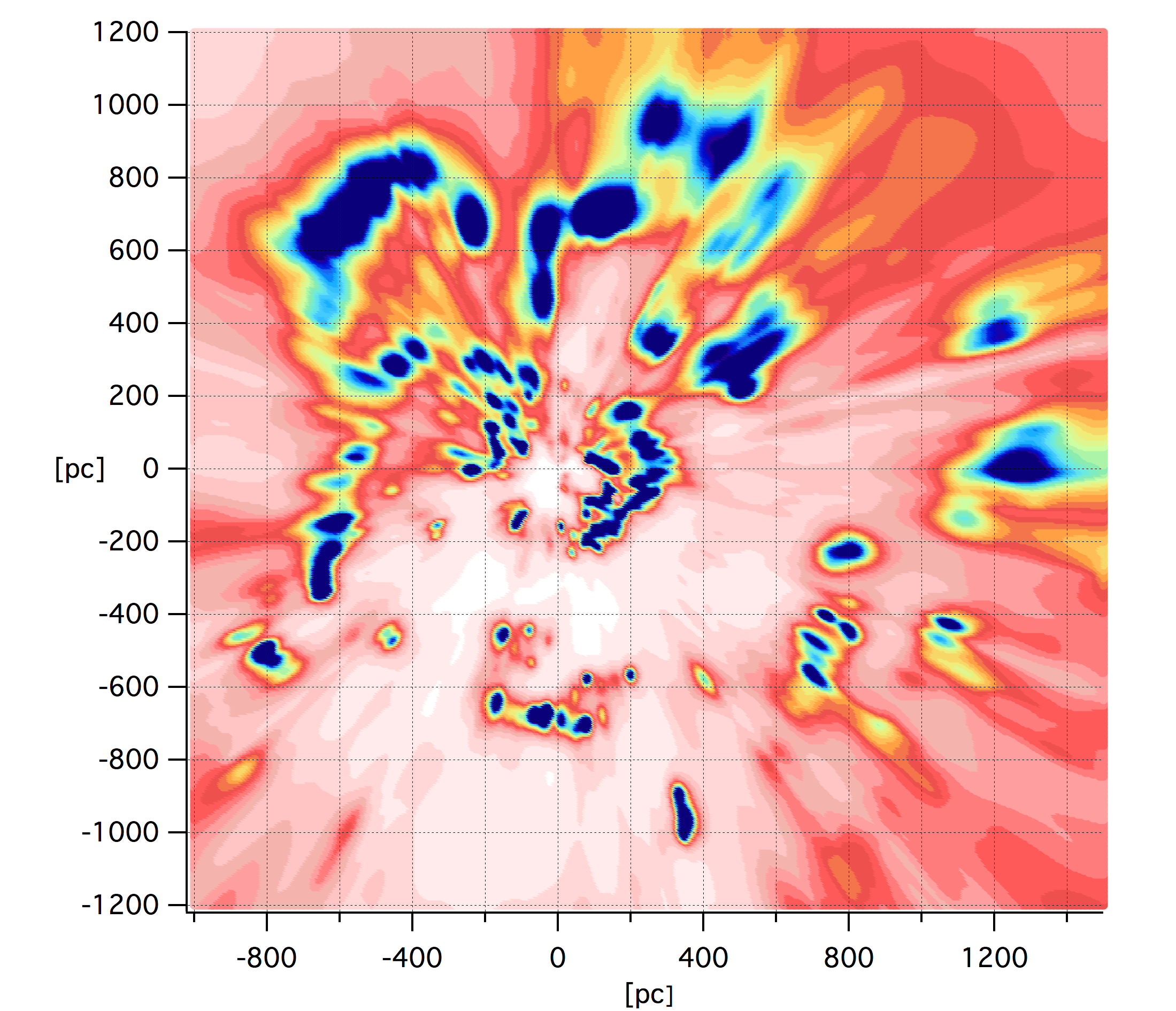}}\\
\caption{Galactic plane cut in the 3D opacity distribution resulting from the four successive inversions: 
a) A distribution: color excess data alone and Hipparcos or photometric distances.
b) B distribution: same as A except for the replacement of 80\% of the initial distances with Gaia-DR1 (TGAS) values. For the 20\% of stars without Gaia distance we kept the previous value. 
c) C distribution: same as B except for the addition of DIB-based color excess estimates for stars with a Gaia-DR1 
parallax. 
d) D distribution: same as C except for the use of a Pan-STARRS based prior distribution instead of a homogeneous distribution.
Color scale, direction of Galactic center, and Sun position are the same as in Fig \ref{priorps}}.
\label{allgalplanes}
\end{figure*}

\begin{figure*}
\centering
\includegraphics[width=10cm]{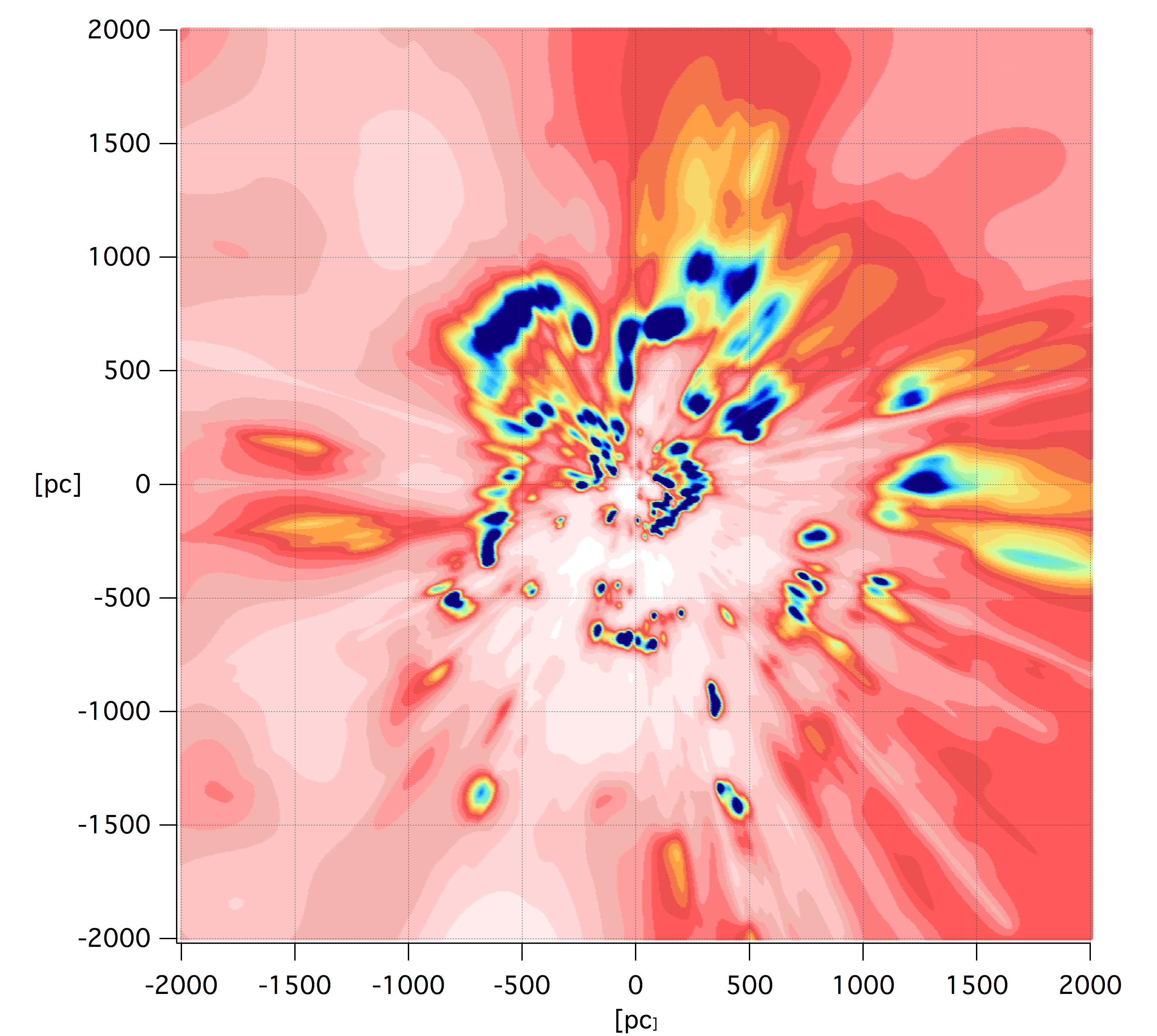}
\includegraphics[width=10cm]{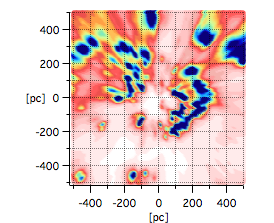}
\caption{Planar cut along the Galactic plane in the inverted 3D distribution D shown in Fig \ref{allgalplanes}, panel d. The color scale is same as in Fig \ref{priorps}.
 Top: the entire 4000pc $\times$ 4000pc computed area. Bottom: Zoom in a 1000 pc $\times$ 1000 pc area around the Sun.} 
\label{FinalPlane}
\end{figure*}

\begin{figure*}
\centering
\includegraphics[width=12cm]{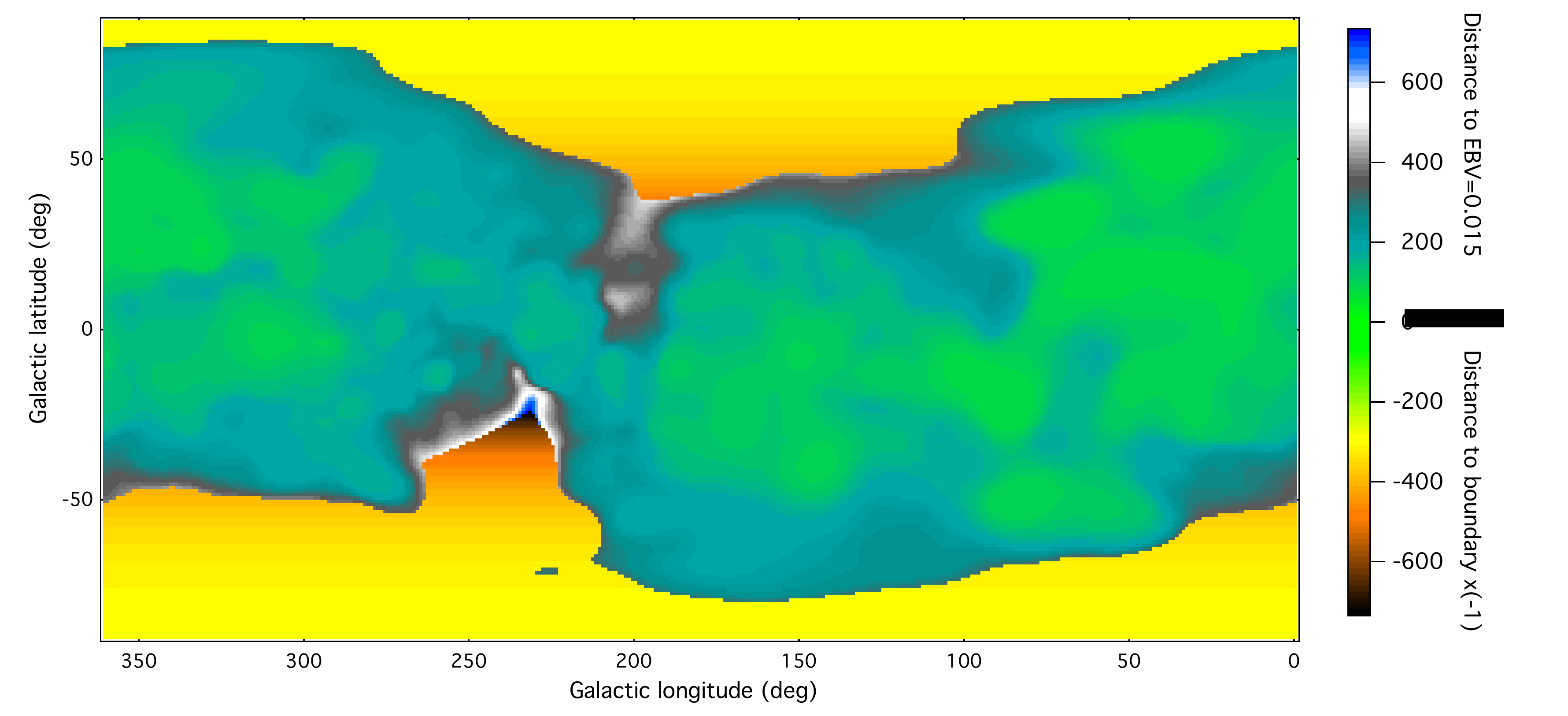}
\caption{Local cavity boundaries: the distance at which a color excess of 0.015 mag is reached is plotted as a function of rectangular Galactic coordinates. In very weakly reddened directions for which 0.015 mag is not reached within our computation domain, the distance to the domain boundary is plotted after conversion to negative values. This allows us to distinguish between the two cases well. The figure illustrates well the local warp of gas associated with the Gould belt, i.e., more matter at positive latitudes in the Galactic center hemisphere and at negative latitudes in the anticenter hemisphere. It also reveals very clearly the \textit{local bubble} openings to the halo at about 210-240$\fdeg$ longitude.} 
\label{MAP_EBV_LOWRED}
\end{figure*}

\begin{figure*}
\centering
\includegraphics[width=12cm]{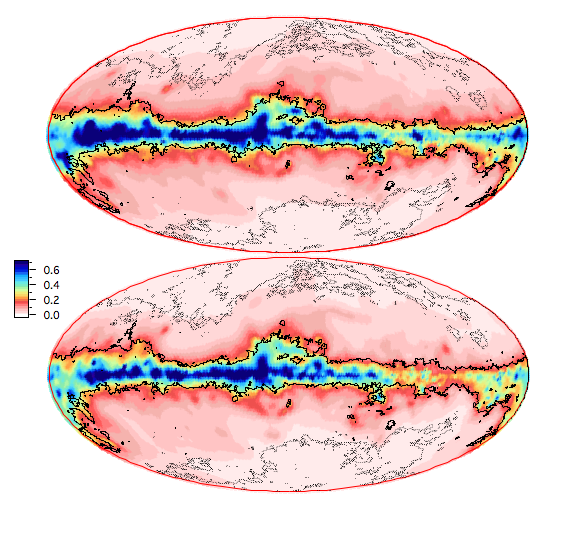}
\caption{Full-sky 2D map of opacity distribution integrated from the Sun to the boundaries of our computational domain (abs(Z)=300pc, abs(X)=2kpc, abs(Y)=2kpc). Top: LVV results are shown (distribution A). Bottom: the final inversion with APOGEE data and the Pan-STARRS based prior distribution are shown (distribution D).  
 Isocontours from \cite{Schlegel98} for E(B-V) = 0.025 and E(B-V)=0.32 are superimposed as dotted and continuous black lines. 
 The map is in Aitoff Galactic coordinates. Longitudes decrease from +180\fdeg{} to -180\fdeg{} from left to right. The color scale refers to E(B-V), linear, in units of mag.} 
\label{MAPLIMIT}
\end{figure*}

\begin{figure*}
\centering
\includegraphics[width=10cm]{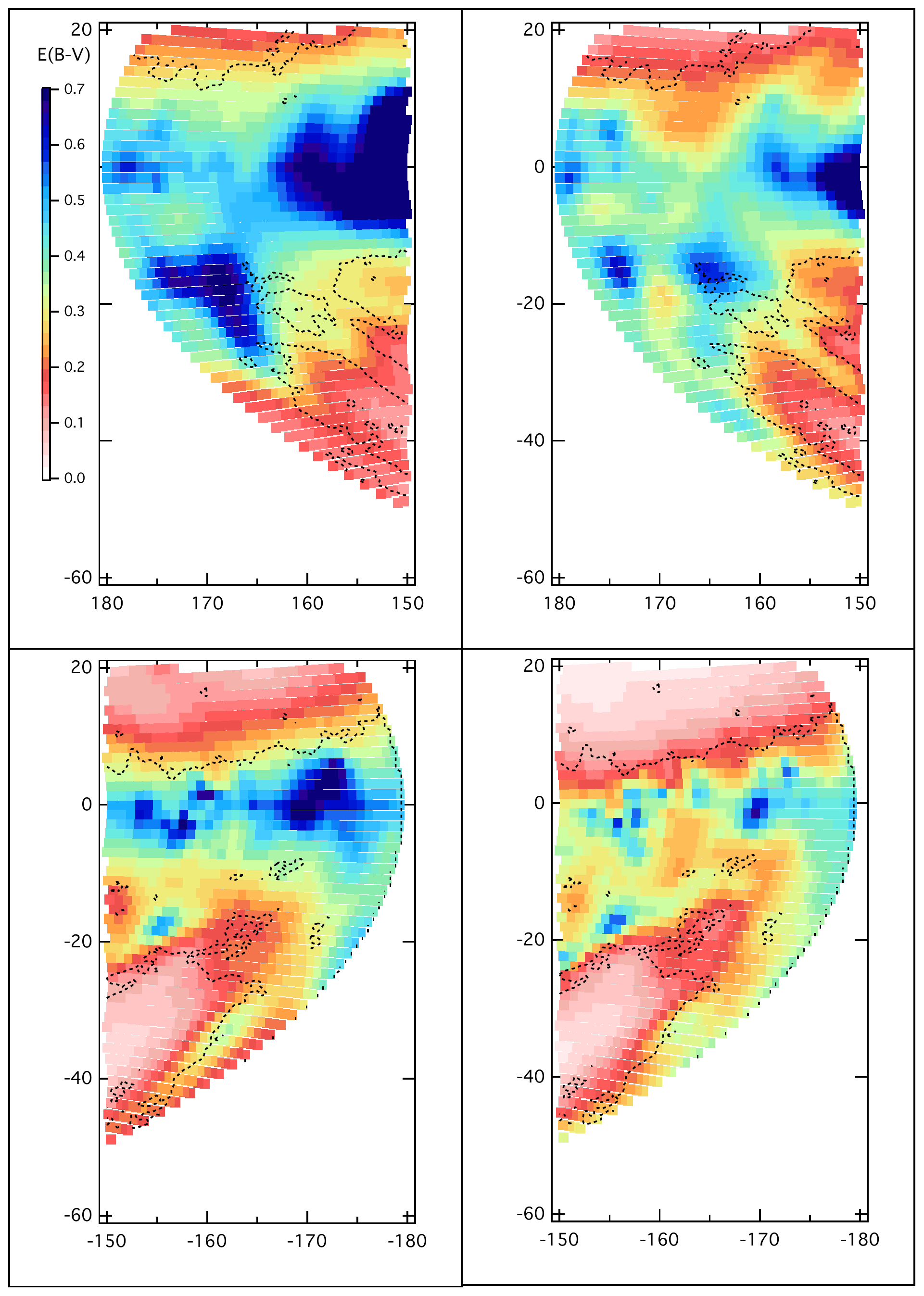}
\caption{Same as Fig. \ref{MAPLIMIT} for limited regions in the anticenter area. Left: LVV data are shown. Right: final inversion is shown. Isocontours from \cite{Schlegel98}  for E(B-V) = 0.025 are superimposed as dashed black lines.} 
\label{MAPLIMIT_AC}
\end{figure*}

\section{Inversion of datasets}\label{inversions}\label{allinversions}

% {Now we want present you the different inversion results, made with different dataset. 
%We produced four 3D maps, see Fig \ref{allgalplanes}, which we recall in the text as A, B, C and D opacity distributions. A ans B opacity maps and have the same targets but different distances, C adds to B dataset the APOGEE targets, and D uses the \prr{} prior.}

\subsection{Inverting pre- and post-Gaia individual color excess data}\label{inversold}

In order to test the influence of the new Gaia parallax distances we inverted the reddening dataset described in section \ref{dataold} twice with exactly the same method, criteria, and 
parameters, but with different target distances. 
The first inversion (A) used the initial Hipparcos+photometric distances and errors as in LVV and the second (B) used Gaia-TGAS distances and associated errors when available, 
or unchanged values in the absence of Gaia result. 
Gaia DR1 parallaxes are about 20 times more numerous than Hipparcos parallaxes and about three to four times more accurate. 
These characteristics will evolve with the next releases, which will contain many more stars and reduced parallax uncertainties. As in LVV the prior dust opacity is a plane-parallel homogenous distribution 
decreasing exponentially from the \textbf{Galactic plane} with a scale height of 200 pc. We also similarly retained for the inversion targets closer to the \textbf{Galactic plane} than 1000 pc and less distant than 2601 pc with relative errors on 
the distance smaller than 33\%. The number of targets retained for the first inversion was 22467 and among these 5106 were assigned their Hipparcos distance, and the rest were photometric. 
In the second inversion 23444 targets were retained, with about 80\% having a Gaia parallax.
  As the Gaia-DR1 uncertainties were 3-4 times smaller than Hipparcos uncertainties, the number of  targets 
excluded for uncertain distances was significantly reduced. Importantly, the fraction of photometric distances decreased by a factor of four and Gaia distances started to be largely dominant.

The values of the $\chi^2$ per target are computed from the differences between the observed color excesses  and the color excesses computed by integration of the model 3D distribution 
along the target lines of sight.  These values are reported in Table \ref{tabchi2} for both the prior and final distribution. 
The value of $\chi^2$ for the prior distribution is significantly higher for the second inversion, reflecting the fact that error bars entering its computation are on average smaller because of the reduced uncertainties on Gaia-DR1 distances, especially by comparison with previous uncertainties on the photometric distances. 
On the contrary, the final $\chi^2$ is lower despite reduced error bars, demonstrating that the found distribution is more compatible with the new distances than with the previous distances. 

Panels a and b (top left and right) in Figure \ref{allgalplanes} show planar cuts in the two inverted opacity distributions A and B. The plane is parallel to the Galactic plane and contains the Sun. 
 The pre-Gaia inverted distribution is not exactly the same as in LVV because of the removal of some targets and the introduction of the new method to treat outliers (see section \ref{classicalinv}). 
The difference between the two pre- and post-Gaia distributions is visible: on average the structures are more compact. Some elongated structures diminished owing to the reduction of distance uncertainties, as in the third and fourth quadrants (bottom left and right); some others appeared, which may be due to contradictions between new Gaia distances and the remaining photometric distances. The latter regions will have to be carefully studied once all targets benefit from Gaia parallaxes. The Local Bubble keeps its global shape, but looks slightly more complex owing to the higher compactness of structures. 
At larger distance the most conspicuous difference appears in the first quadrant, in particular in the Cygnus Rift region. 
At l$\simeq$70\fdeg the Rift is located significantly farther away in the new inversion, at about 800 pc instead of $\simeq$600 pc, interestingly now in agreement with the \cite{Green15} distance assignment. We knew that this was a region of strong oscillations in the first inversion; these oscillations have now disappeared. 
In the second quadrant there are substantial changes and we start to sketch a ring of clouds, more specifically the very large structure located from 100 to 400 pc is  {decomposed into two smaller structures.}

\begin{table*}
\centering                                  % used for centering table
\begin{tabular}{l l l l}        % centered columns (4 columns)
DISTRIBUTION & DATA & INITIAL $\chi^2$  & FINAL $\chi^2$  \\
\hline                                   %inserts single line
 {A} & LVV catalog, Hipparcos+phot & 5.361 &  1.100\\ %nuovo
 {B} & LVV catalog, Gaia-DR1+phot &  5.810 &  1.085\\ %il nuovo
 {C }& LVV catalog + \apo{}, Gaia-DR1+phot & 5.080 & 1.023\\% nuovo
 {D} & LVV catalog + \apo{}, Gaia-DR1+phot ,  \prr{} prior&  3.521  & 1.006\\ %nuovo
\end{tabular}
\caption{ Evolution of the prior and final $\chi^2$ per line of sight in the four inversions.}             
\label{tabchi2}
\end{table*}

\begin{figure}
\centering
\subfloat[][B opacity distribution]
{\includegraphics[width=.45\textwidth]{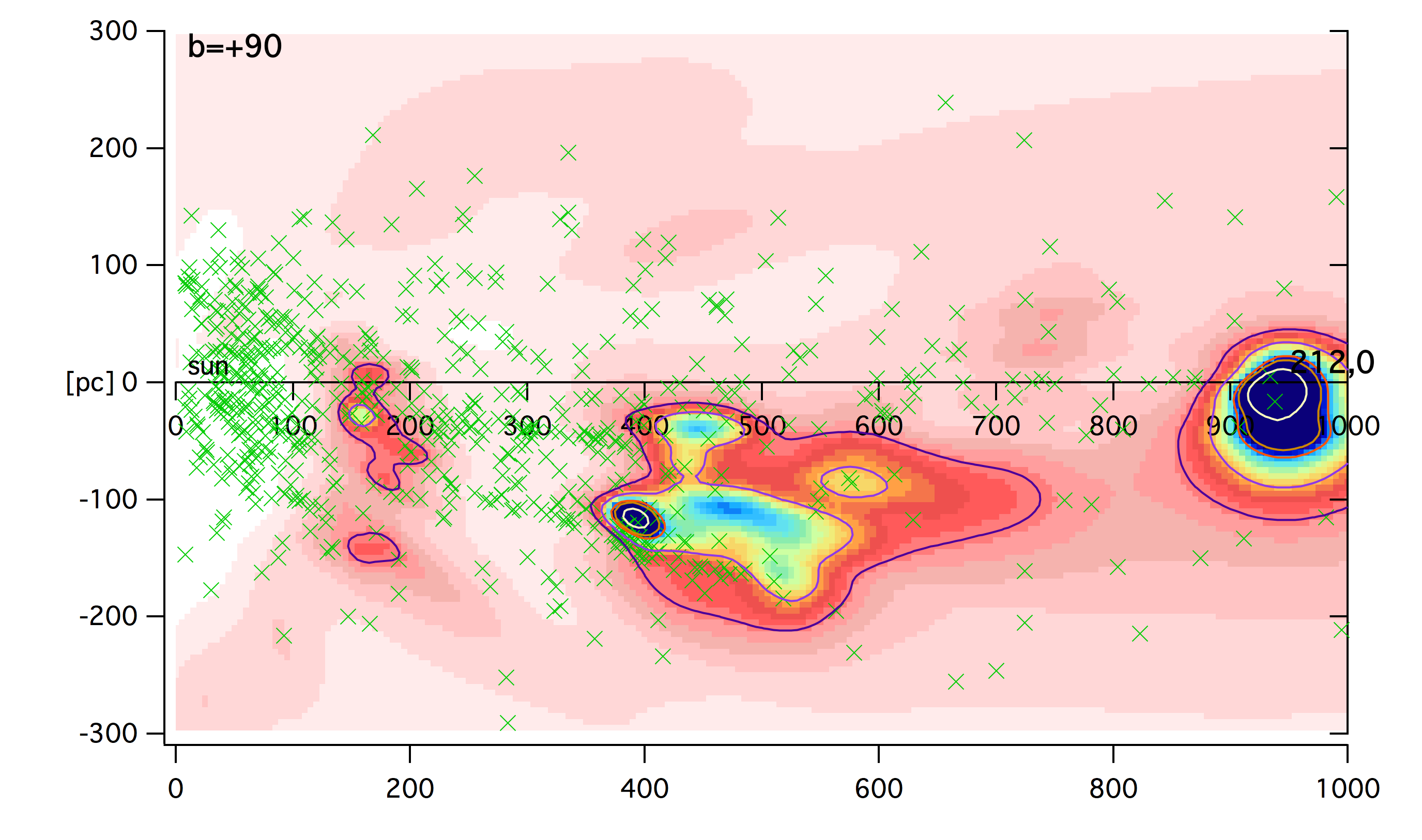}}\\
\subfloat[][C opacity distribution]
{\includegraphics[width=.45\textwidth]{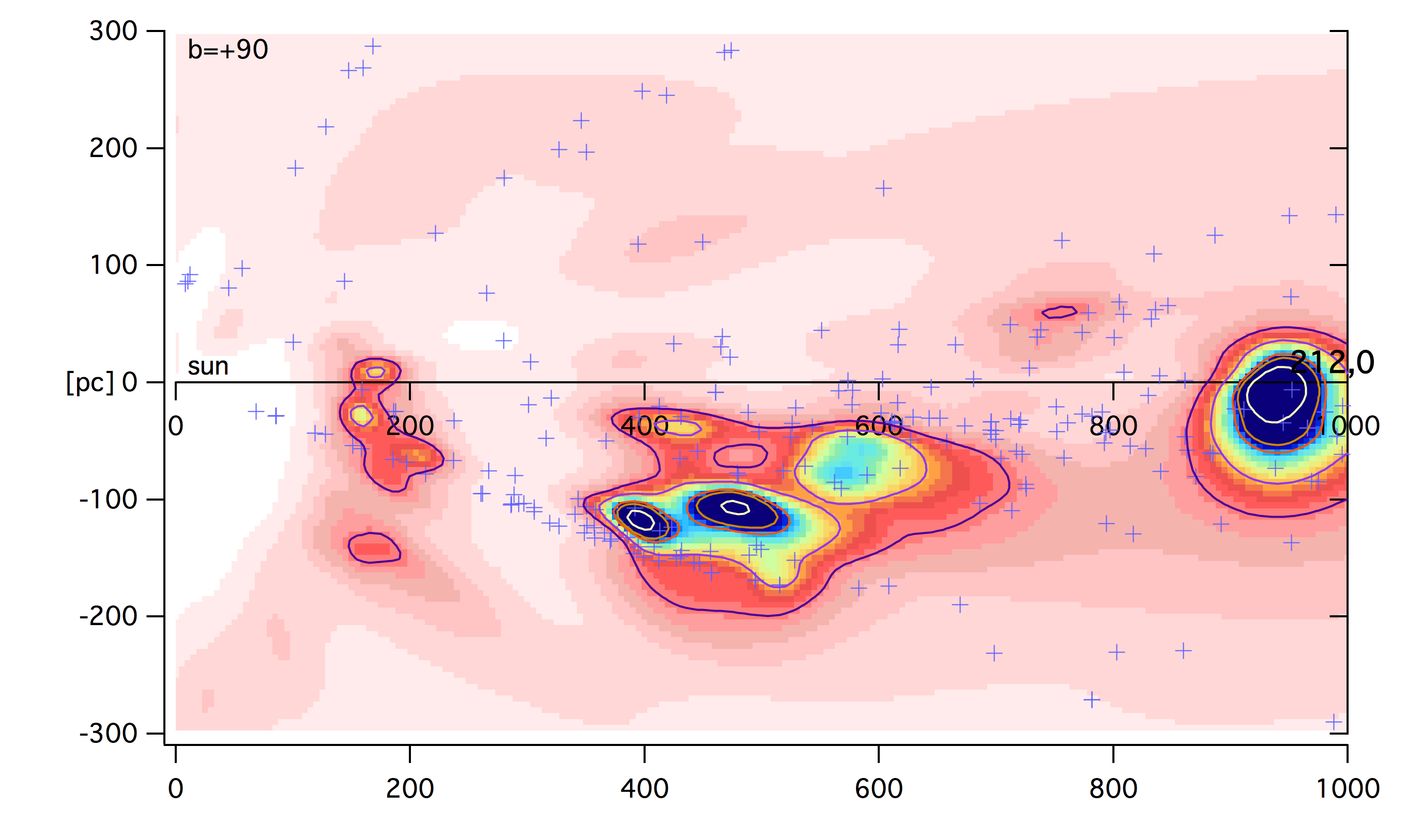}}\\
\subfloat[][D opacity distribution]
{\includegraphics[width=.45\textwidth]{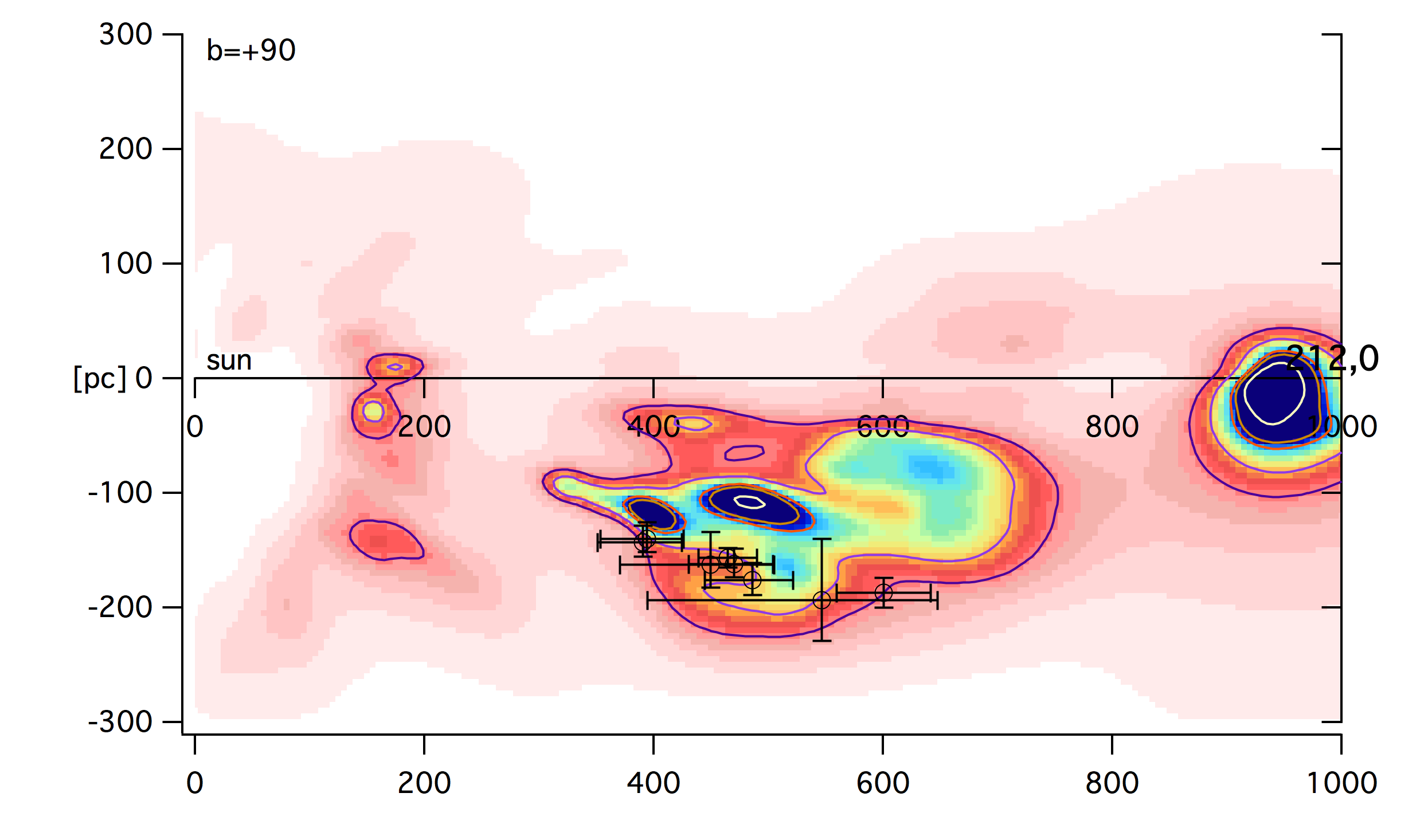}}
\caption{Comparison between the three B-C-D distributions (see Sect. \ref{allinversions} and Fig. \ref{allgalplanes}) for a vertical plane along l= 212\fdeg. The X-axis is within the Galactic plane, the Y-axis is toward the North Galactic pole direction.  The color scale is the same as in Fig. \ref{priorps}. Iso-contours are superimposed for $\rho_0=$ = 0.0002,
0.00035, 0.00076, 0.001, and 0.002 mag.pc$^{-1}$.  The green crosses  {in top panel} are the positions of the targets from the LVV catalog in a slice of 10 degrees around the  {considered vertical} plane, violet crosses  {in middle panel} 
are the targets from the \apo{} catalog, and the
black circles  {in bottom panel} are the molecular clouds positions and associated uncertainties based on the catalog of \cite{Schlafly14}.}
\label{cut212}
\end{figure}

\begin{figure}
\centering
\subfloat[][B opacity distribution]
{\includegraphics[width=.45\textwidth]{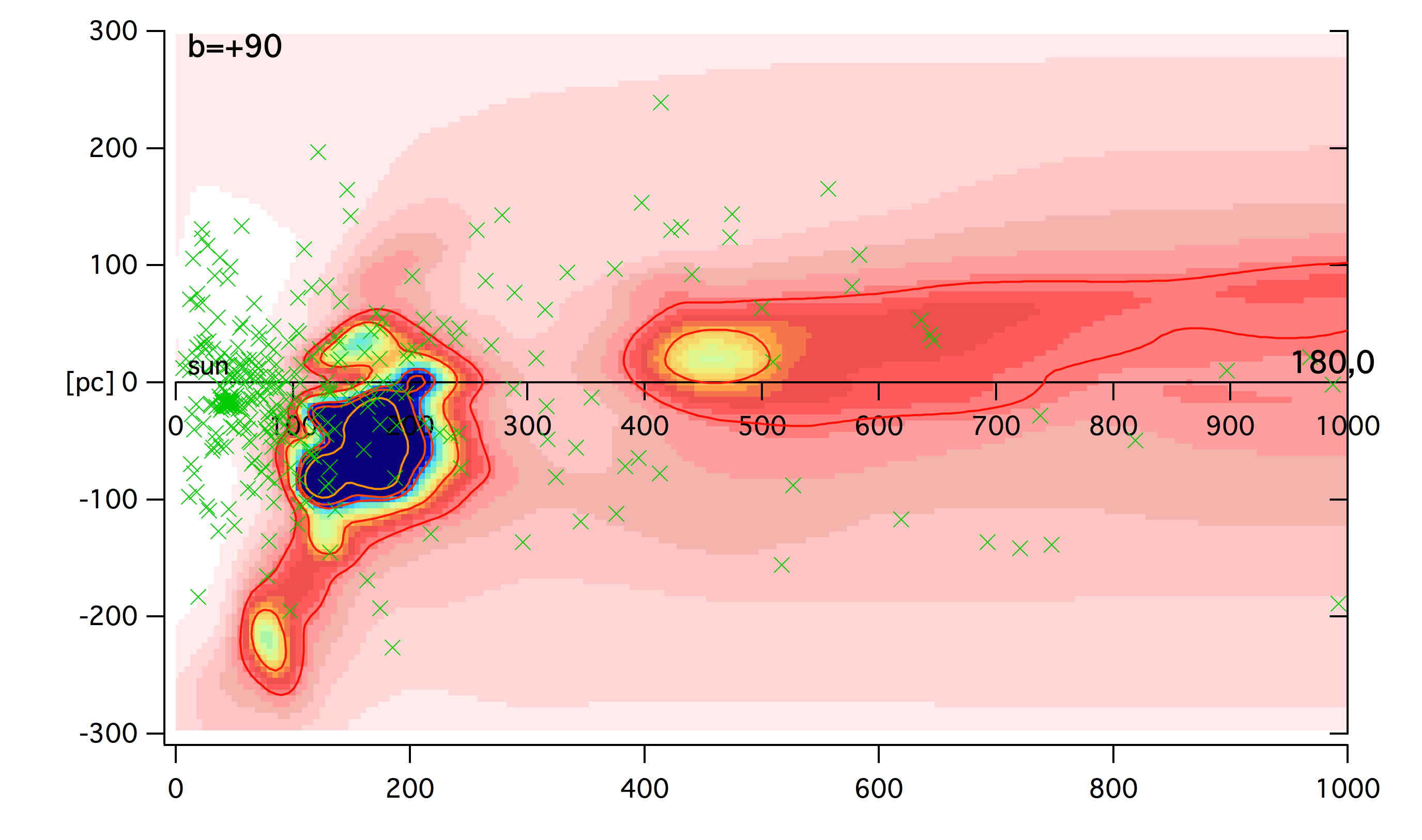}}\\
\subfloat[][C opacity distribution]
{\includegraphics[width=.45\textwidth]{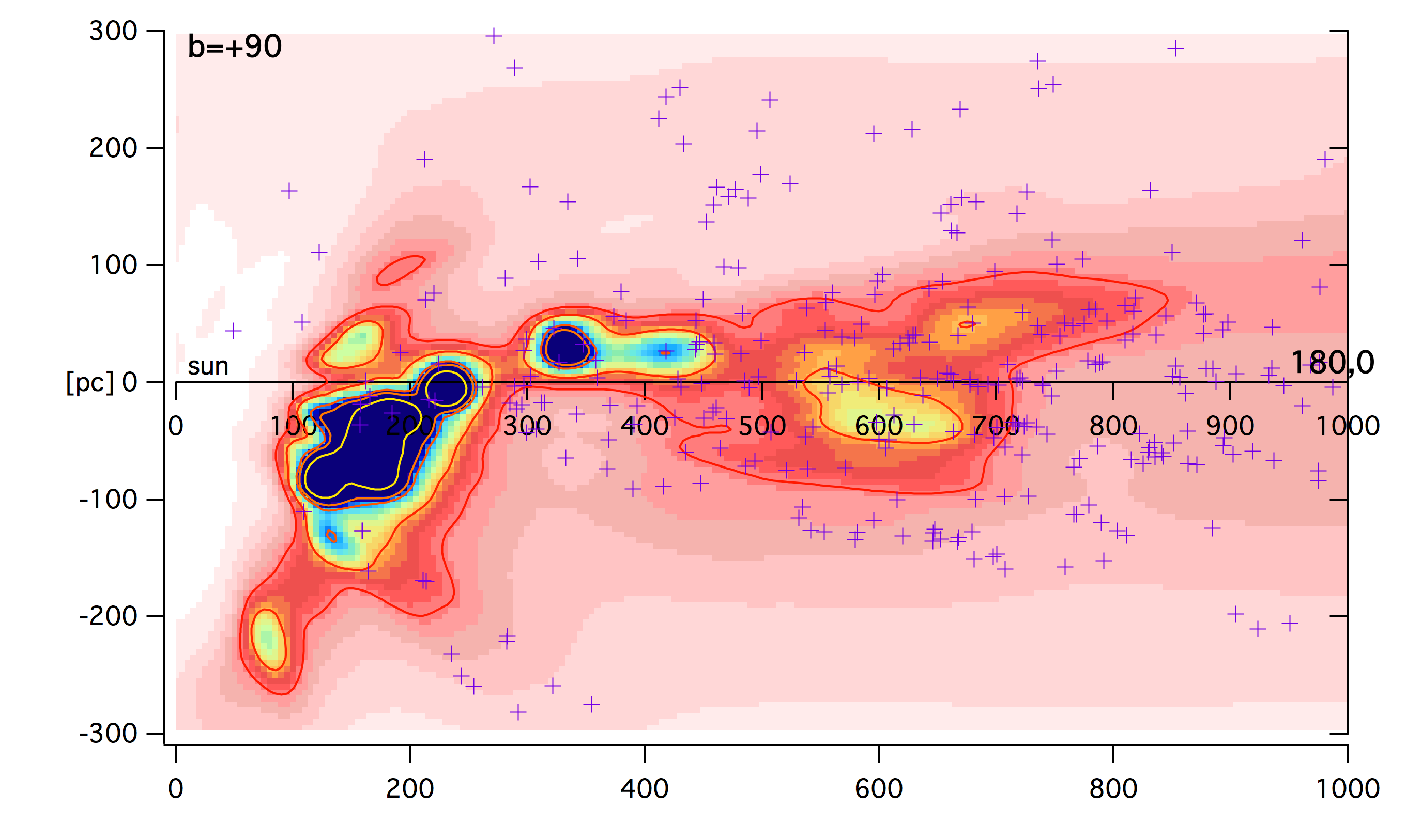}}\\
\subfloat[][D opacity distribution]
{\includegraphics[width=.45\textwidth]{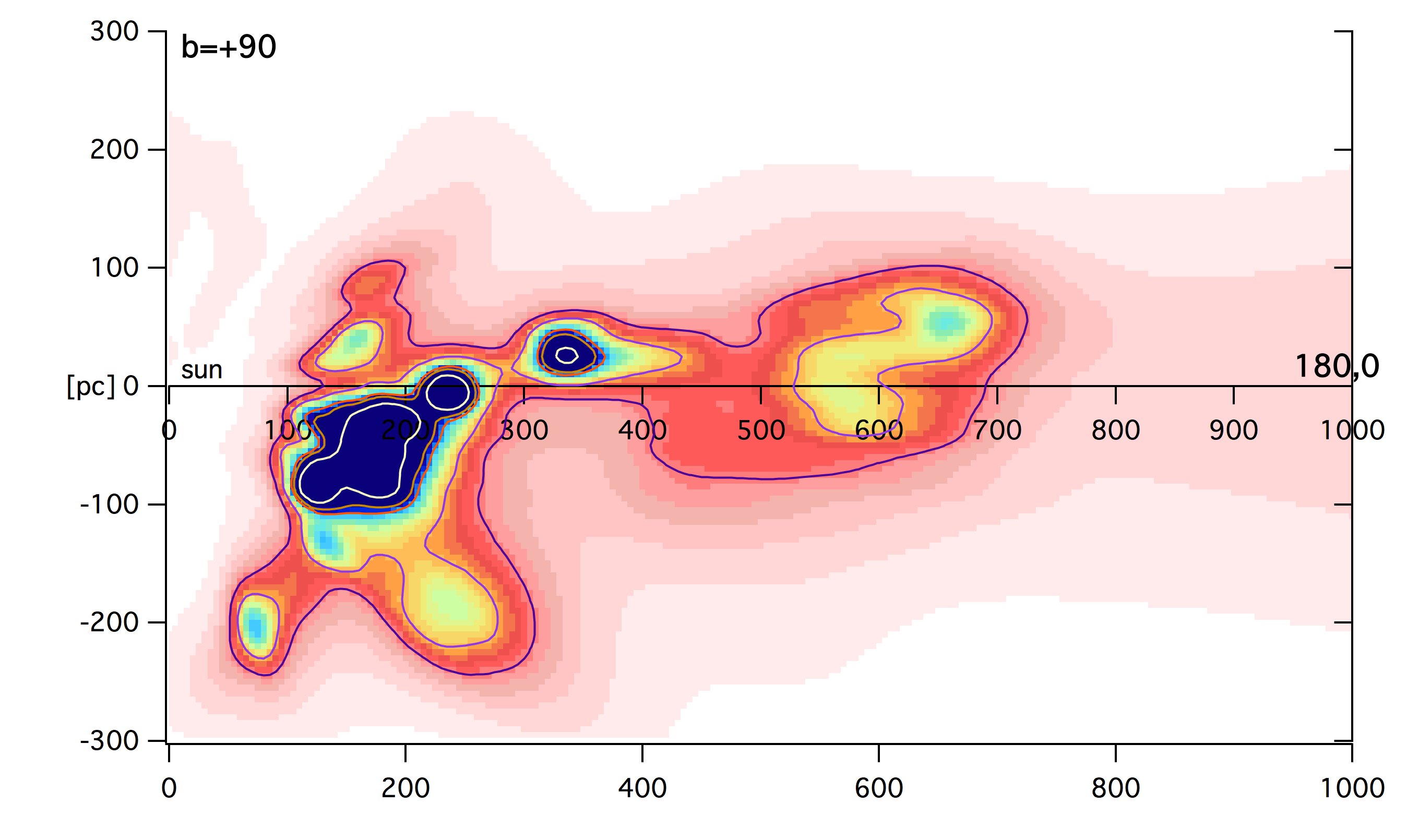}}
\caption{Same as fig \ref{cut212} for the vertical plane along l= 180\fdeg (anticenter region). }
\label{l180new}
\end{figure}

\begin{figure}
\centering
\subfloat[][B opacity distribution]
{\includegraphics[width=.5\textwidth]{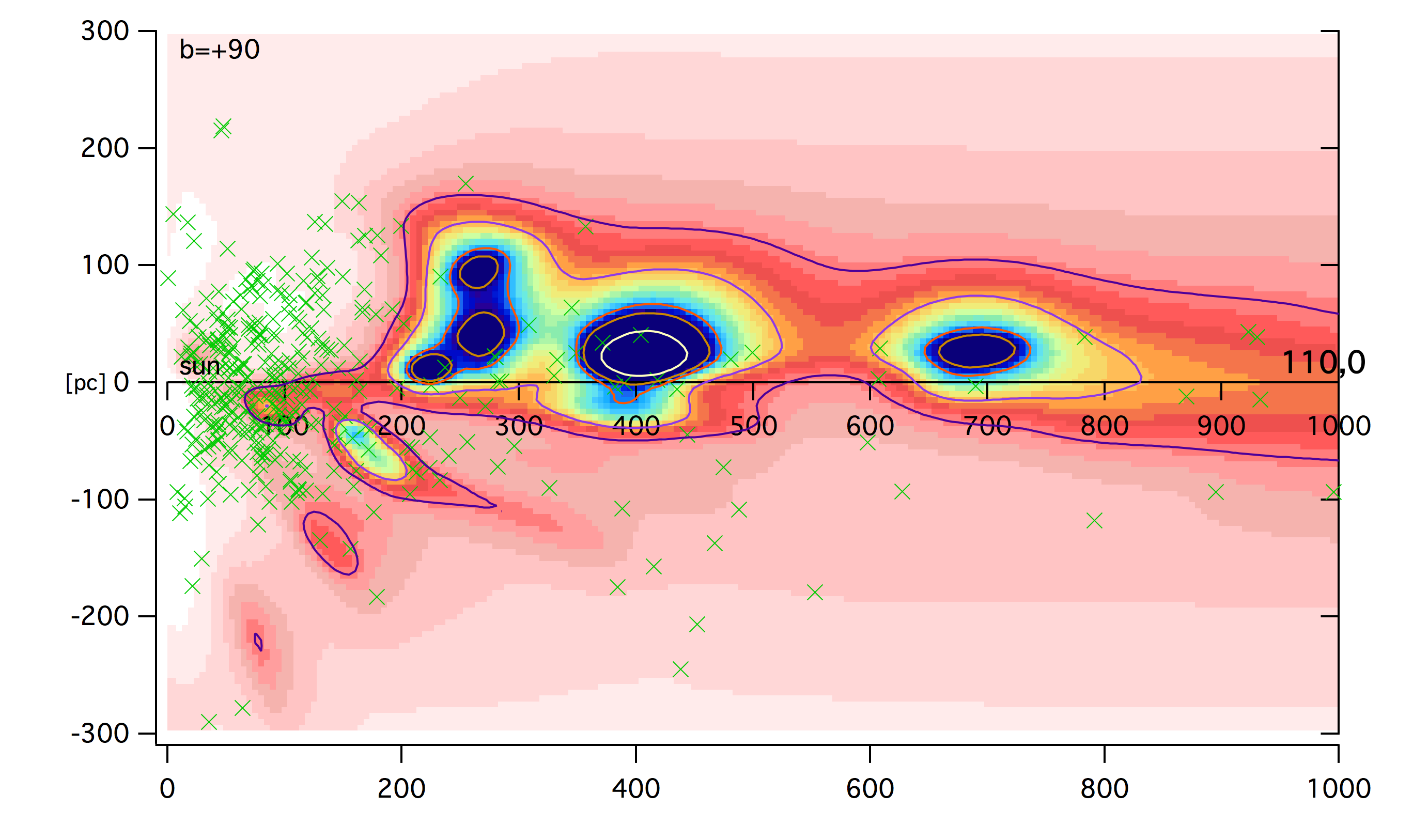}}\\
\subfloat[][C opacity distribution]
{\includegraphics[width=.5\textwidth]{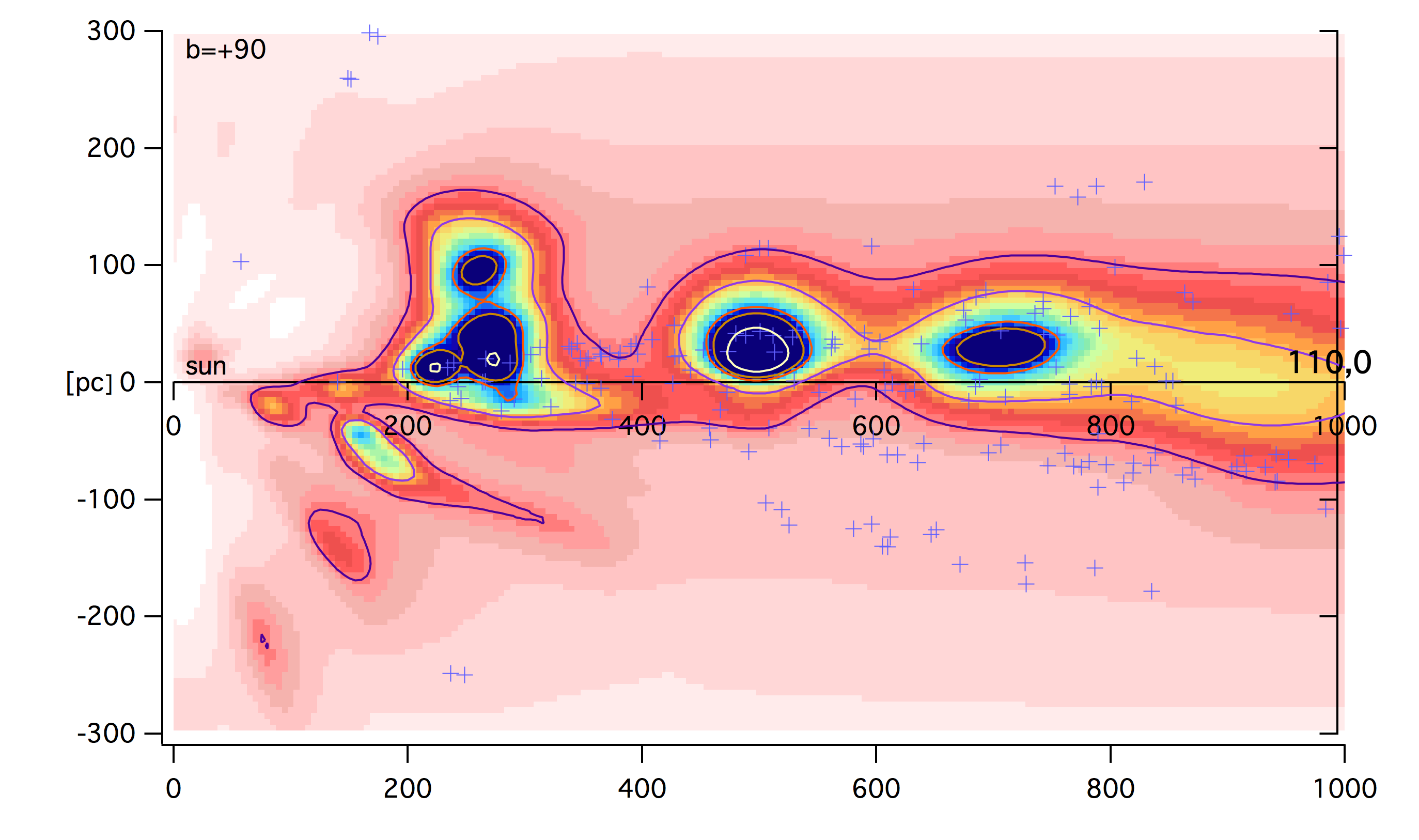}}\\
\subfloat[][D opacity distribution]
{\includegraphics[width=.5\textwidth]{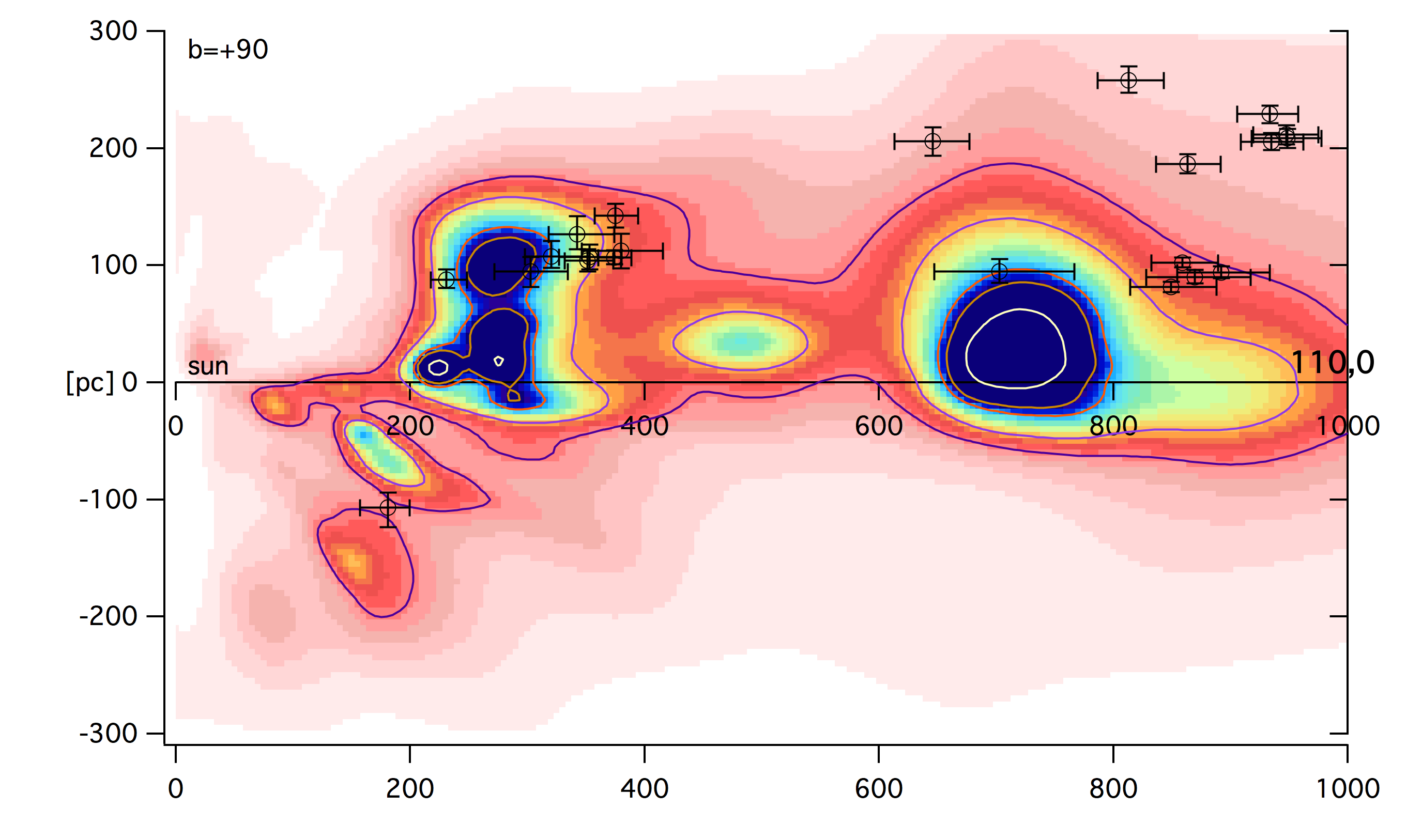}}
\caption{Same as fig \ref{cut212} for the vertical plane along l= 110\fdeg (Cepheus region).}
\label{cut110}
\end{figure}

\begin{figure}
\centering
\subfloat[][B opacity distribution]
{\includegraphics[width=.5\textwidth]{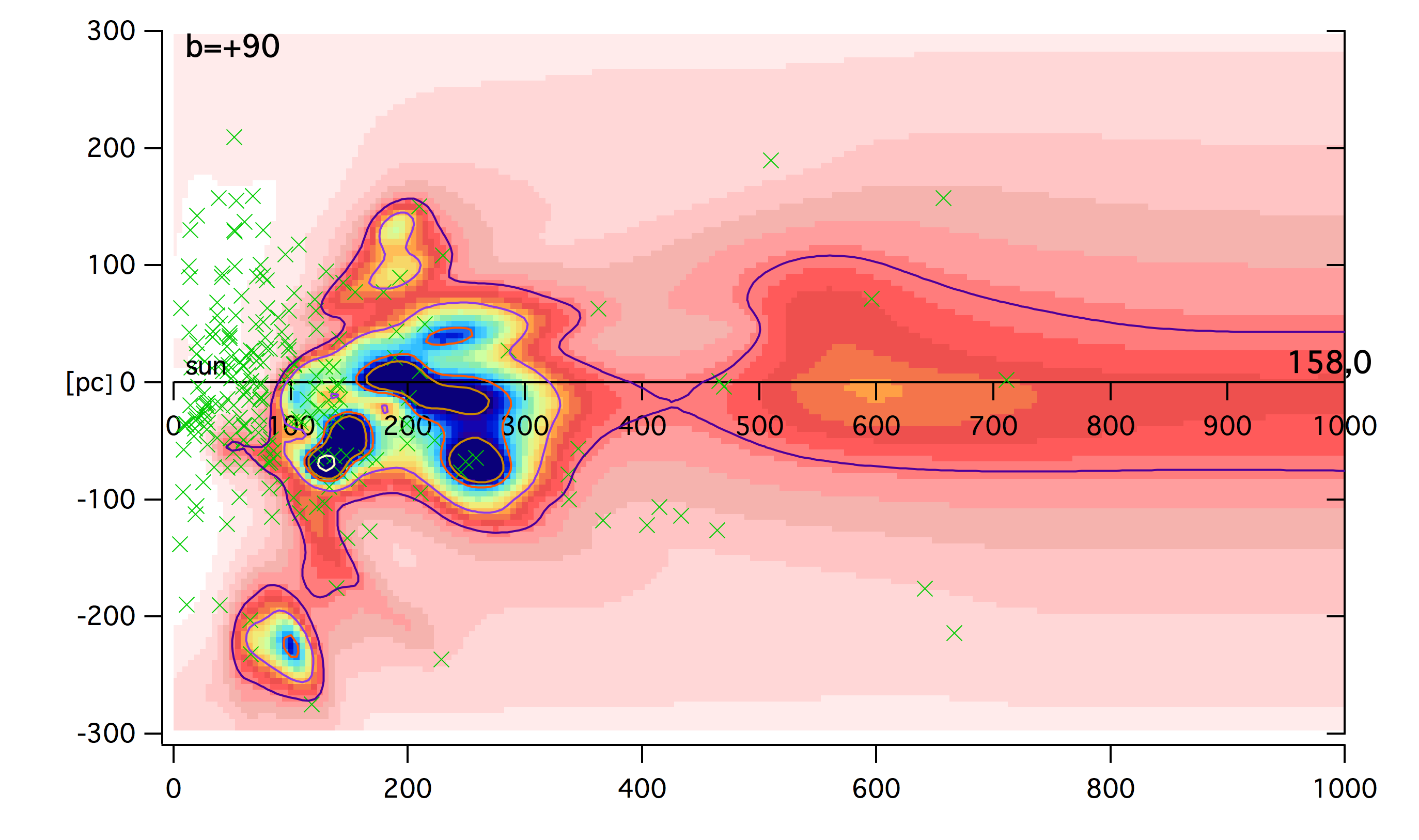}}\\
\subfloat[][C opacity distribution]
{\includegraphics[width=.5\textwidth]{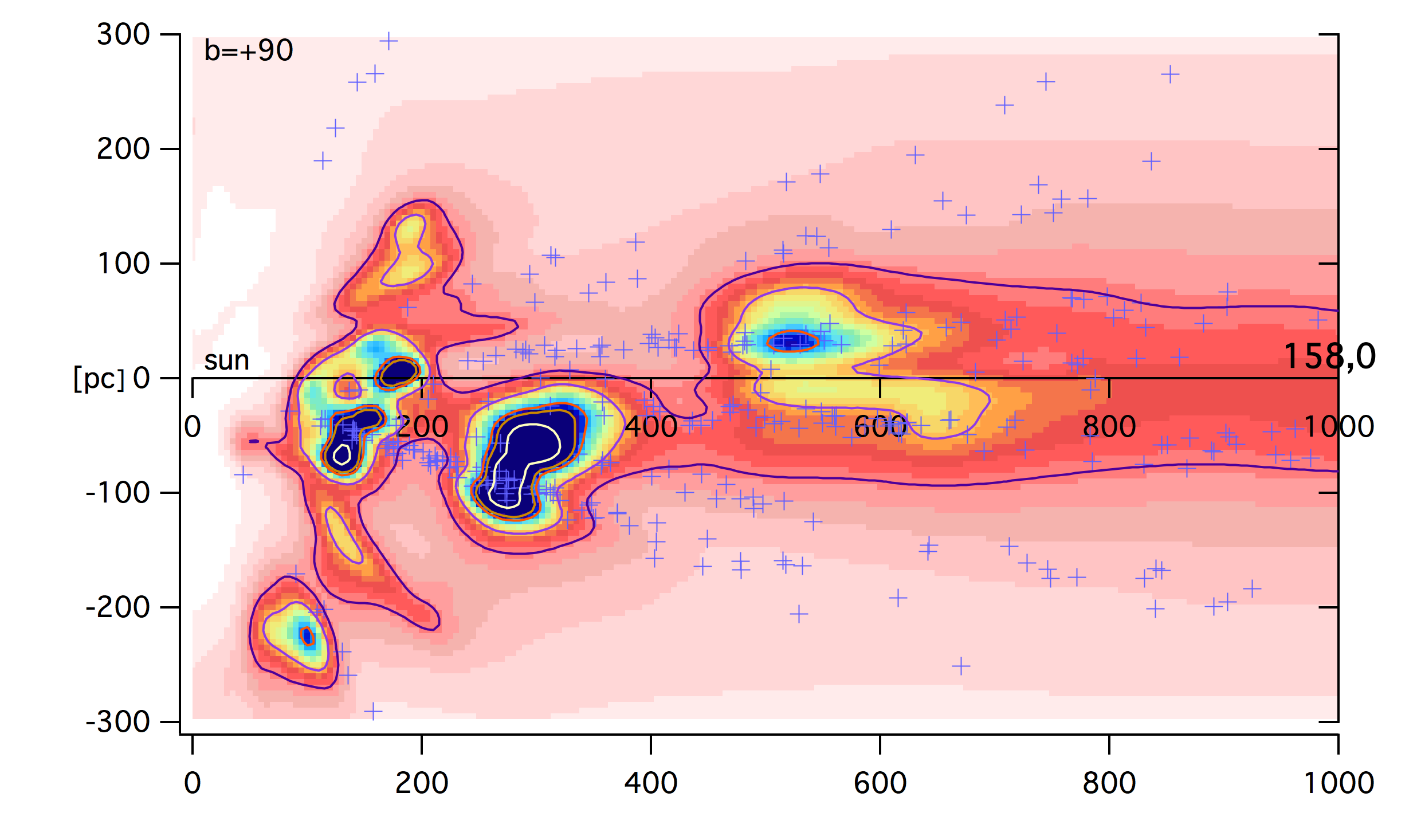}}\\
\subfloat[][D opacity distribution]
{\includegraphics[width=.5\textwidth]{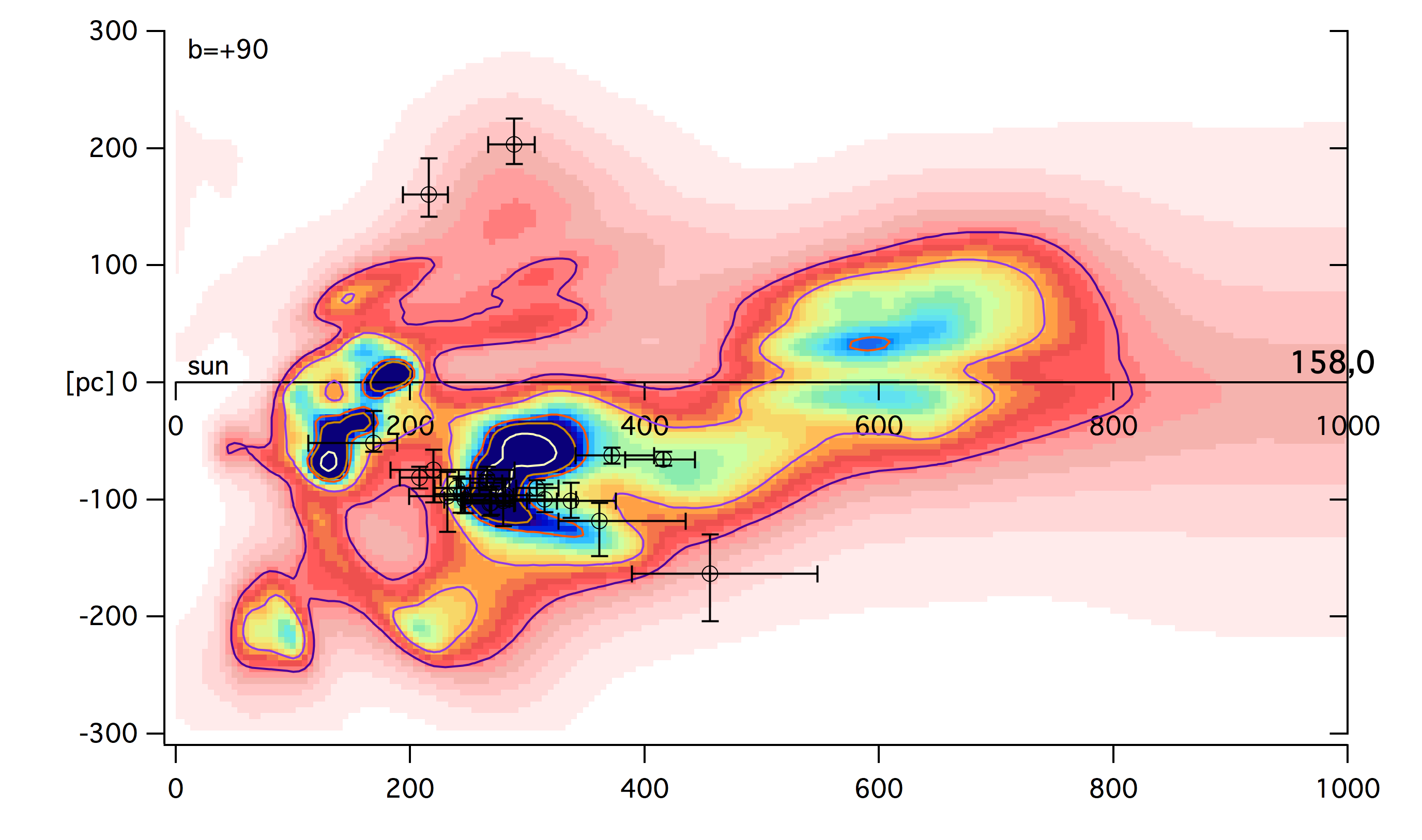}}
\caption{Same as fig \ref{cut212} for the vertical plane along l= 158\fdeg (Perseus region). }
\label{cut158}
\end{figure}

%The first inversion 

%The main part of our data are in the galactic plane, 6379 line of sight has plane altitude between -20 pc and 20 pc - almost 20\% of total data. This plane is the most detailed, here we can observe some difference between the difference maps. 

%Figure \ref{fig:oldplanes} a) is the Galactic plane with the data as in \cite{rosine:oldmap}. The colour scale 
%is in logaritmic scale. Comparig with the same inversion with Figure \ref{fig:oldplanes} b), we observe some changes in the region between l = 60 and l = 90, where the structures are sharper; the fingers of God are less important and the clouds shapes more contained. With more coherent distances, the global inversion is simpler and it gives better $\chi^2$, compare Tab \ref{tab:chi2}. In Fig. \ref{fig:verticalcutO} vertical cut every $30^\circ$.
\subsection{Inversion of combined photometric and DIB-based color excess data}\label{inversapogee}

The same inversion code was applied to the merged catalog of photometric and DIB-based reddening estimates  {(inversion C)}. As can be seen in Table \ref{tabchi2}, the introduction of the \apo{} targets reduces both the initial and final $\chi^2$ per point. The former decrease means that error bars on the DIB-based estimated E(B-V)s are not underestimated. 
The latter decrease strongly suggests that the APOGEE targets do not introduce contradictory data and their inclusion produces a reasonable solution. 
In comparing the new Galactic plane cut in Fig \ref{allgalplanes} and the distribution C at the bottom left {with the previous inversion with photometric data alone (distribution B, top right)}, we can clearly see some changes; these changes are particularly apparent in the second quadrant, where some
 decoupling or fragmentation of large structures into distinct cloud complexes along with new structures at large distances are visible. 
 Since 14 \% of the \apo{} targets are closer than 20 pc from the \textbf{Galactic plane}, 
 are highly concentrated in this quadrant, and are more distant that the LVV  targets, it is not surprising to find most changes there. 
 The fourth quadrant does not change because the \apo{} catalog adds targets observed from the Northern Hemisphere.

%But it's interesting also look at some others planes. For example, we look at the plane and we note that there are important changes
 %in the region l=110,130, slice where we have added 717 line of sight at the previous 672. Fig \ref{fig:cut110}, Fig \ref{fig:cut120}, Fig \ref{fig:cut130}; 526 \apo{} lines of sights have -in 
 %this slice - distance larger than 500 pc, 149 lines were the previous. In the farer part we observe the presence of new 
 %structures, in the inner part the sharpness of several little structures.  
 %a different technique to validate the DIB 1.5273 $\mathrm{\mu m}$ in the single linesight maps as generic matter tracer.
 
%Looking the Fig \ref{allgalplanes}, we observe the general fragmentation in smaller structure for the clouds in the second quadrant, in the 
%closer 500 pc. The addition of data in a second bell outside the Local Bubble define the external structures. 
Since it is the first time that such measurements of different types are used, which could trigger incoherence in the inversion, 
it is mandatory to investigate the observed changes in all areas in order to validate this new technique. 
 Four vertical planar cuts in the 3D opacity map are shown in Fig \ref{cut212} (Orion region), Fig \ref{l180new} (anticenter direction), Fig \ref{cut110} (Cepheus region),  and
Fig \ref{cut158} (Perseus region). The Perseus region is discussed in detail in Sect \ref{PSINV}.
%Now we present different cloud region: Orion, Cepheus and the anti-center. 
%To make this analysis we use will be use hereafter in Sect \ref{PSINV}.
The images show, in the top panel, the B opacity distribution results and, in the middle panel, the C opacity distribution after inclusion of APOGEE data.
In principle, the effect of adding coherent data in poorly covered areas should result in the transformation of wide structures into smaller clumps and/or in the \text{appearance} of new structures, 
especially at large distances in locations where there was initially no targets. 

The Orion region is well suited for comparisons owing to the existence in similar proportions of targets of the two types beyond 150 pc. % {we use Fig \ref{cut212} to compare the inversion results with
%dataset B (top image) and dataset C (middle image)}. 
There are $\simeq$750  lines of sight from the photometric catalog and $\simeq$375 from the \apo{} catalog within the longitude interval l=(208:218$\fdeg$). 
As shown in Fig \ref{cut212},  the two types of targets  are spatially well mixed. 
If the data were not compatible we would observe radially elongated and/or wide structures in the  {middle} map that were absent in the  {top}. 
Instead, there are no significant changes, which demonstrates that the two datasets trace the same structures. 
 {Indeed, the figure shows that the \apo{} data reinforce the dense structures at 500 and 600 pc,
at distances poorly covered by targets with photometric measurements.}

Another region suited to a further check is the anticenter region. At variance with the Orion area, the LVV targets are concentrated in the first 100 pc, while APOGEE introduces many more distant targets. Fig \ref{l180new} shows clearly that this addition allows  {us to map new clouds beyond 300 pc, in particular to transform a unique wide structure at $\simeq$ 450pc into at least three structures located between 300 and 700 pc}, confirming that infrared data such as APOGEE add valuable information in areas where stars are already reddened by close clouds. The structures  at small distances already seen in the first map  (inversion B) remain in the second map but appear more compact with their relative opacities somewhat changed. Because the sightlines to the distant targets cross the nearby clouds, contributing to their mapping, the increased compactness of the foreground structure shows that the new data do not contradict the old data and instead help shape the close clouds. 

We finally consider the Cepheus region in the second quadrant (Fig \ref{cut110}).
 There are several clouds distributed in the same region of the sky, which is a very good test for the inversions. 
Here, the number of targets in the longitude interval  l=(105$\fdeg$;115$\fdeg$) increases from 390 to 664.  
It can be seen that the large structure at distances between 200 and 300 pc and at heights above the \textbf{Galactic plane} between 0 and 150 pc is now more clearly decomposed into three cores and an additional fourth cloud just below the \textbf{Galactic plane} at $\simeq$ 300 pc.
This last structure below the \textbf{Galactic plane} had apparently been erroneously associated spatially with a more distant and wide structure at 500 pc. 
As a result of Gaia distances and additional targets, its location has now been better defined. Indeed, the final distribution D (bottom panel of Fig \ref{cut110} and section \ref{PSINV}) 
confirms this new location. Behind those structures a hole as well as two different more distant clouds are revealed near the \textbf{Galactic plane}. The large structure located at about 400 pc in 
distribution A, which was derived from only a few targets, is now much better defined and significantly displaced farther away at about 500 pc.  
This again shows that using infrared data enables the mapping of distributed clouds and holes between them.
 
 \subsection{Inversion of merged data using the Pan-STARRS-based prior 3D distribution} \label{PSINV}

We performed an additional inversion  {(inversion D)} of the set of LVV+ APOGEE targets, this time replacing  the plane-parallel homogeneous prior with the large-scale distribution described in Sect. \ref{hierarchical}, based on \cite{Green15} results. This allowed us to test our new capacity of using an arbitrary 3D prior distribution. In the new prior, the smallest structures allowed are 100 pc wide and the kernel sizes for the largest structures are 800 pc. 
%\begin{enumerate}
%\item A priori knowledge of the MIS: an analytical model with a homogeneous extinction in the galactic plane which decreases exponentially according to the height on the galactic plane
%\item Use of \prr{} data to constrain the opacity of the interstellar medium at large scales: development of a more detailed map than the analytical model described in 1)
%\item Use of individual lines of sight and Gaia parallaxes to constrain the opacity on finer scales from the map obtained in 2) which is used as a prior.
%\end{enumerate}
%Obviously, nothing forces us to stop at step 3): the map obtained in step 3) can itself be refined with new data.

%Our last inversion is identical to the previous one except that instead of a homogeneous prior distribution we use the distribution described in section \label{hierarchical}. Fig \ref{priorps} shows the Galactic plan for this prior based on \cite{Green15} Pan-STARRS-based reddening curves. Since  \prr{} data cover the Northern sky down to $\delta \approx $30 $\deg$, regions below this declination 
%keep the homogenous prior. The smaller allowed structures are 100 pc and the larger ones 800 pc. 

%For the last inversion, the \emph{prior} for the extinction per parsec changes from a function of the plan distance to a map got from the \prr{} observation at large scale. 
Table \ref{tabchi2} shows that using the new prior has the effect of reducing the initial $\chi^2$ per point significantly and decreasing the final $\chi^2$  too. 
This means that the data for individual sightlines are, not surprisingly, in better agreement with this more realistic prior. 
A comparison of panels c and d in Fig. \ref{allgalplanes} shows the differences introduced by using the new prior  {in the Galactic plane}. As expected, the main variations are at large 
distance, especially beyond 1 kpc, while closer the changes are very small. 
% In the Orion region, as seen  in Fig \ref{cut212} or in other vertical cuts in the distribution, \
 The same effects are seen out of the \textbf{Galactic plane}, as shown in  Fig \ref{cut212} (Orion), Fig \ref{l180new} (anticenter), Fig \ref{cut110} (Cepheus), and Fig \ref{cut158} (Perseus), where the 
 bottom panels show the D opacity distribution. There are significant changes far from the Sun and negligible changes closer to the Sun, where the target density is high enough to suppress 
 the influence of the prior. In all cases, it is interesting to see how the nearby structures connect with the distant very low-resolution Pan-STARRS based distribution. This two-step process
  allows large-scale and smaller-scale structures to coexist in a map, while avoiding an overinterpretation of the data. Other vertical planes drawn every 30\fdeg Galactic longitude are shown 
  in Appendix. These vertical planes give us a general idea of the nearby ISM distribution out of the \textbf{Galactic plane}.

Fig \ref{FinalPlane}  {(upper panel)} shows the dust opacity in the Galactic plane for the entire computational area 4000pc x 4000 pc. As can be seen in the Figure, at large distances there are co-existing compact structures with the minimum size allowed by the kernels and large-scale structures that mainly result from the Pan-STARRS prior. The small structures however should be considered with caution, as is discussed in the next section.

Bottom panel of Fig \ref{FinalPlane} shows a zoom in the 1000pc x 1000pc central area. It shows numerous clouds and cavities and that the shape of the so-called Local Bubble is far from 
simple. It is well known that the interstellar matter within the first hundreds of parsecs is very complex and characterized by the series of clouds forming the Gould belt,
 a wide structure tilted w.r.t. the \textbf{Galactic plane} by about 20$\fdeg$ (e.g., \cite{Perrot03}). The origin of this structure is still a matter of debate (see, e.g., \cite{Lall47tuc} for a recent discussion and 
 suggested origin). Two \textit{chimneys} connect the Local Bubble to the northern and southern halos forming a hole whose axis is nearly perpendicular to the Gould belt plane;  {this peculiar 
 structure is illustrated in Fig \ref{MAP_EBV_LOWRED}. The figure} shows for all directions the line-of-sight length that corresponds to an integrated color excess E(B-V)=0.015 magnitude. 
 When the distance required  {to reach this color excess} is larger than the distance to the boundary  of  the computational volume, such as toward high Galactic latitudes, we used this 
 distance multiplied by -1. The figure shows very clearly the local warp of IS matter as well as the \textit{chimneys} connecting to the halo.

 Fig. \ref{MAPLIMIT} shows 2D color excesses from inversions A (top map) and D (bottom map), which are computed by integrating local opacities from the Sun to the boundaries of our computational domain; the distance to the \textbf{Galactic plane} is smaller than 300 pc and distances along the X and Y axes are smaller than 2,000 pc. Iso-contours from \cite{Schlegel98} for E(B-V) = 0.025 mag and E(B-V)=0.32 mag are drawn for comparison.  
Compared to our previous inversion (A, at top) the structures are more compact for inversion D (at bottom), showing that the data are more consistent and the resolution is improved thanks to Gaia distances and additional data. Fig. \ref{MAPLIMIT_AC} is a zoom in on these 2D maps of A and D integrated color excesses, this time in a limited region of the sky close to the anticenter direction.
 Compared to our previous inversion, D leads to better agreement with the \cite{Schlegel98} contours at low reddening, demonstrating that the inclusion of DIB-based data has improved the detection and location of the structures.

%We will discuss the resolution/distribution problem in Section \ref{error}. We want to stress now how using a non homogeneous prior the inversion changes at farer distances. 

%The anticenter come from this inversion is also largely compatible with the previous, because the structure at 300 pc appeared adding the \apo{} data is in the same position and the 
%farer  

%For the longitude 158, the Perseus nebula has the closer part with the same shape and the distances are largely compatible, but the hole which separated the Perseus nebula from the farer structure in the North galactic hemisphere are now linked from a "bridge". In this case the prior got from observation, despite it is only fro large structures improves the 
%global quality.

%It is possible have a bias with the prior at large structure which pretend to have connected structures where we have only closer ones, but no data to separated them; it is not the case
%in the 180 longitude because of the \apo{} targets distribution, as in Fig{l180new}. 

\section{Uncertainties and comparison with other 3D measurements}

\subsection{Resolution and errors} \label{error}

Uncertainties on cloud distances and dust columns are strongly dependent on the distance to the Sun due to the combined effect of target space density decrease and target distance uncertainty increase. Additionally, uncertainties vary from one region to the other since databases have strong biases. Theoretically the errors may be calculated for each point in space along with the Bayesian inversion, but this kind of evaluation is very expensive in computation time (\cite{SaleMag14}). On the other hand less precise error estimates for the distance to the structures and the amount of integrated reddening may be obtained by means of simpler approaches. We present these approaches here. 

\begin{figure*}
\centering
\includegraphics[width=.3\textwidth]{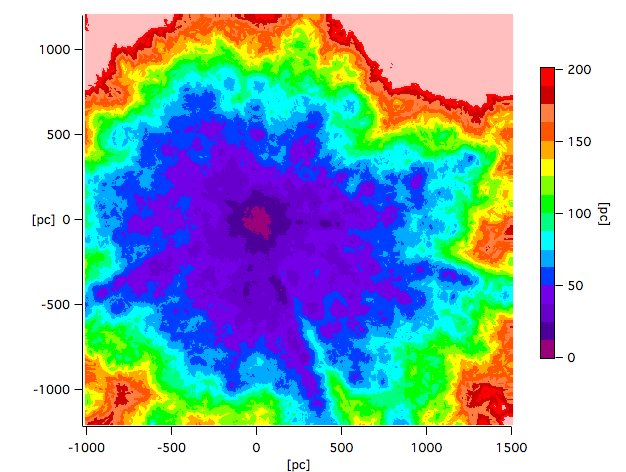}
\includegraphics[width=.3\textwidth]{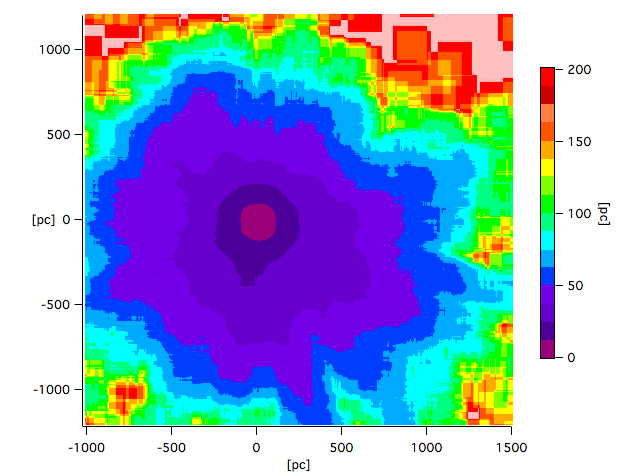}
\caption{Error in the distance due to the limited target space density,  in the Galactic plane,  as estimated from the radii of 
the smallest spheres that contain at least 10 targets (left figure) or from the target density in a cube of side 200 pc (right). The pale pink color corresponds to errors larger than 200 pc.}
\label{errorplan}
\end{figure*}

In the frame of a first simple approach we assumed that the error in distance varies as  {the achievable resolution}, i.e., depends on the target space
 density. We choose as an estimate of the error at each point the size of the  {smallest} sphere around this point that contains at least 10 targets. Another simpler and 
computationally faster approach is to calculate again in each point the target density $\rho$ (in pc$^{-3}$) in a cube of a given size centered on this point and use (1/$\rho$)$^{1/3}$ 
as the error estimate. This second technique is less precise because we keep the same volume everywhere in an arbitrary way. The results for the Galactic plane are 
shown in Fig \ref{errorplan}. The results from both methods are not markedly different and as expected there is a strong increase of the distance uncertainty from the 
center to the peripheral regions. The error can be compared with a similar estimate by LVV: it can be seen that the addition of the \apo{} targets improves the resolution 
in a number of regions from the first to third quadrant. The errors on cloud localization owing to uncertainties on the target distances are  {limited by the use 
of our threshold (see Sect \ref{inversold}).}

\begin{figure*}
\centering
\includegraphics[width=.3\textwidth]{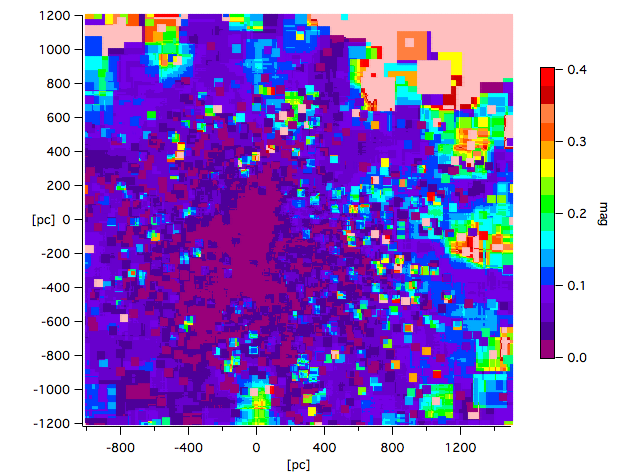}
\caption{Estimated error on the integrated color excess for the Galactic plane computed as the mean difference between data and model in a restricted volume (see Sect. \ref{error}). The pale pink color 
corresponds to errors larger than 0.4 mag. }
\label{errorebv}
\end{figure*}

As already mentioned in LVV, estimating the error on the local dust opacity is meaningless because by principle the clouds have a minimum size and as a consequence 
the dust is distributed in larger volumes in the model compared to the volumes of the actual, clumpy clouds. 
 {This spreading of the opacity in volumes that are wider than the volumes of the actual clouds introduces systematic underestimates of the opacities and integrated reddenings in the cloud cores. 
As a result, this systematic error is expected to be small for sightlines that do not intercept small, dense cores, 
but may reach high values in highly reddened regions in case the reddening is generated by one or a few small structures. We come back to this point in Section \ref{conclusion}. 
On the other hand, it is possible to give rough estimates of the errors on 
the reddening (integrated opacity) that are linked to the dataset, independently of the errors due to the assumed cloud minimum size. 
To do this, for every point in our 3D map we selected all targets 
contained in a cube of specific size around the considered point, and for those targets we calculated the mean difference  between the data and model color excess.} 
Fig \ref{errorebv} shows the results for the \textbf{Galactic plane} with a cube of 50x50x50 pc$^3$; in case the cube contains less than 10 targets we use a volume of 200x200x200 pc$^3$. 
Uncertainties on distances and color excess will be available on our website and displayed interactively.

%One approach is to use the error on the distances discussed above, integrate the opacity from the Sun to the limits of the uncertainty "sphere", derive the minimum and maximum values of the integrals and use this interval as an error bar. 
%VOIR ce qui est sur le site et finalement choisi)XXX
\subsection{Molecular clouds catalogs}\label{molcloudcomp}

%Our map is not a density map, but extinction per parsec. But there is a link between the matter presence and the reddening and excitation, it depends from the dust grains dimension, composition, and others factors. We could not use our map as a density map, but we know that the dust and cold gas produce the effect that we map. 
\cite{Schlafly14} produced a catalog of molecular cloud distances based on \prr{} photometric data. 
Most of the clouds are out of the \textbf{Galactic plane}, where the number of structures distributed along the sightline is relatively small (see Fig \ref{Targets1and2}). 
These authors have mapped the clouds cores with a very high angular resolution, while our inversion produces much wider structures, however it is interesting to use this catalog for a comparison of cloud locations.

In Fig. \ref{cut212}, \ref{cut158}, and the bottom panel of Fig. \ref{cut110},  we superimposed the opacity distribution resulting from our  {D} inversion and the position of the 
cloud cores identified by  \cite{Schlafly14} with associated error bars on their  {positions}. For the Orion region ( {galactic longitude = 212$\fdeg$}; Fig. \ref{cut212}), 
these authors found a sequence of clouds distributed from 400 pc to 600 pc, which are distances similar to the structures appearing after the inversion. 
In the case of Perseus ( {galactic longitude = 158$\fdeg$}; Fig \ref{cut158}), our map also indicates the same distances. 
In the case of the Cepheus region ({galactic longitude =110$\fdeg$}; Fig. \ref{cut110}) the inversion produces two large, opaque structures above the {Galactic plane}, 
the first between 250-300 pc and the second between 700-800 pc.  {Two groups of high latitude molecular clouds have also been mapped in this region by \cite{Schlafly14}. The closest group is in reasonable agreement with our mapped structures, while the second is at somewhat larger distances.}  {However, in this case our targets are very scarce at high latitude (see upper and middle
panels in Fig \ref{cut110}) and more data are needed to allow a comparison}. 
%On the other hand, they derive a series of more distant clouds (between 800 and 1000 pc)  {which} are absent from our maps. 
%As the locations of the targets show (fig \ref{cut110} top-middle), this is due to the lack of distant targets, and we need more data to be able to perform comparisons about these distant features.

%\begin{figure*}
%\centering
%\includegraphics[width=.5\textwidth]{cut_l108o}
%\includegraphics[width=.5\textwidth]{cut_l108oa2}
%\includegraphics[width=13cm]{cut_l108oa2jl5}
%\caption{Vertical cut at longitude 108$\deg$. Black circles are cloud positions as identified by \cite{Schlafly14}.}
%\label{schlafly108}
%\end{figure*}

\subsection{Soft X-ray emitting cavities as signatures of nearby recent supernovae}

\begin{figure*}
\centering
\subfloat[][]
{\includegraphics[width=.35\textwidth]{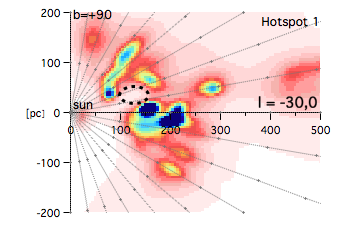}}
\subfloat[][]
{\includegraphics[width=.35\textwidth]{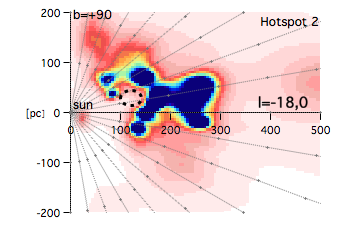}}\\
\subfloat[][]
{\includegraphics[width=.35\textwidth]{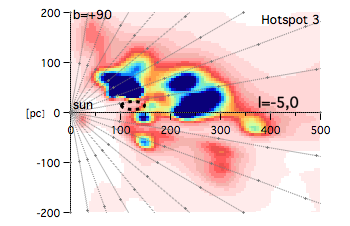}}
\subfloat[][]
{\includegraphics[width=.35\textwidth]{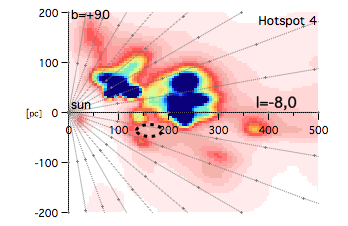}}

\caption{Four vertical planar cuts in the 3D opacity maps in planes containing the ROSAT 3/4 keV X-ray bright spots labeled 1 to 4 in \cite{Puspitarini14} and here as a) to d); the black dashed ellipses show the likely locations of the corresponding SNRs; the latitudinal grid is every 10 $\deg$. }
\label{rosat}
\end{figure*}

Following the work of \cite{Puspitarini14} we searched for cavities in our 3D maps, 
which can be counterparts of supernova remnants (SNRs) or massive stellar winds.
Such cavities are detected through their soft X-ray emission in ROSAT maps. 
Fig. \ref{rosat} shows four vertical cuts in the 3D distribution along meridian lines crossing the four ROSAT 3/4 keV bright spots labeled 1 to 4 by \cite{Puspitarini14}. 
The coordinates of those X-ray spots are l,b = (-30$\fdeg$, +14$\fdeg$ ), (-18$\fdeg$, +18$\fdeg$), (-5$\fdeg$, +9$\fdeg$ ), and (-8$\fdeg$, -10$\fdeg$).
In the four cases, we see that at the latitude of the bright spot centers there are cavities surrounded by dense clouds, which is the typical configuration for moderately old SNRs. 
Moreover, these cavities are not separated from the Sun by opaque clouds, which is a condition for their detection in soft X-rays, 
since million K gas soft X-rays  photons are easily absorbed; more precisely, the cavities must have foreground extinction that is lower than E(B-V)=0.1 to be easily detected. 
Compared to the previous search using LVV maps, the cavities are now better defined. Interestingly, {the two cavities seen in the maps at l,b,d=(-18$\fdeg$,+18$\fdeg$,125 pc) and (-30$\fdeg$,+14$\fdeg$,125 pc) are within 3$\fdeg$ and 7$\fdeg$, respectively, from} the directions of the estimated sites of the two most recent SNRs {(l,b = (-33$\fdeg$,+11$\fdeg$) and (-17$\fdeg$, +25$\fdeg$), \cite{Breitschwerdt16}), and at similar distances}. These authors have used the signature of $^{60}$Fe in deep-sea crusts as a sign of recent (about 2.2 million years ago) explosions of supernovae in the solar neighborhood, and, based on proper motions and ages of the young stellar groups, they computed the most probable present locations of the latest explosions. 
Given the uncertainties associated with their statistical modeling, the small differences between our cavities {1 and 2} and their results is negligible, and these identifications reinforce their model and the $^{60}$Fe measurements and interpretations.
%Starting from the first inversion using the old catalogs with Gaia distances, we observe in direction l = -40 a new cavity and we suspect it come from 
%a supernova occurred in the past. 
%As in \citep{Breitschwerdt16}, they propose two supernova-events in direction l = -33 and l = -17.
%Their techniques are statistical and model depending, so it is quite interesting have some observation in this regions. 
%In this directions there are some little bubble in our maps, and they could strengthen the supernova evidences. 
%The comparison with the X-spot reveals four little bubble in the ROSAT directions. This bubbles are open to our direction, so the 
%X-ray are not absorbed and we can observe them; in Fig \ref{rosat} we compare, as in \citep{Puspitarini14}.The results do not change with
%Gaia distances, therefore the hypothesis that this bubbles are connected with the soft-Xray emissions is strengthened.
 
\section{Conclusions and perspectives}\label{conclusion}

We updated and improved our previous 3D maps of the local interstellar dust, based on the Bayesian inversion of photometry-based color excess measurements, in several ways. Our goal is to analyze the influence of the use of new Gaia parallaxes and to test two additional techniques  {aimed at providing more accurate 3D maps}. First, the merging of photometric color excesses with color excesses estimated from diffuse interstellar band absorptions and, second, the use of a non-homogeneous large-scale prior distribution deduced from massive photometric surveys. The first technique increases the number of sightlines available for the inversion and, in turn, the  {resolution}. The second  {work} aims at testing hierarchical techniques to be used for Gaia photometric data in
the future .
\begin{itemize}
\item{We replaced $\simeq$  5,100 Hipparcos and $\simeq$  17,400 photometric distance measurements with $\simeq$  18,700 new Gaia-DR1 available parallactic distances and $\simeq$ only  {4,700} remaining photometric distances. In the resulting maps some elongated structures have been suppressed, which demonstrates that such features were artefacts owing to the lower accuracy of former distances. Significant changes are also seen in specific areas, in particular the Cygnus Rift.}
\item{We extracted by means of a dedicated profile-fitting technique a series of $\simeq$ 5,000 equivalent widths of 15273 \AA\ diffuse interstellar bands  from SDSS/APOGEE high-resolution spectra of targets possessing a Gaia parallax and converted their equivalent widths into E(B-V) color excesses.  We allowed for an additional 50\% uncertainty to take into account the absence of a strict proportionality between the DIB  {EW} and the color excess E(B-V).  We then performed a new inversion using the combination of the previous reddening data and the DIB-based reddening estimates. In general the APOGEE targets are more distant than the previous stars, which results in more extended maps. We do not find  {any contradiction} between previous and new maps in regions where both types of targets are present in similar numbers.  {Instead, a number of structures that were broad and had low opacities are assigned more complex shapes and denser cores, which suggests} that the DIB-based color excess estimates can enter the Bayesian inversion and be used to get better constraints on the cloud locations. This has important consequences because all spectroscopic surveys provide DIB measurements that may help locate the {interstellar} clouds, especially if all targets have precise parallax measurements.}
\item{We used a subsample of \cite{Green15} Pan-STARRS reddening estimates to build a low-resolution 3D map by means of a new hierarchical Bayesian interpolation method. This map is used as a prior during the Bayesian inversion of the individual data instead of a homogeneous distribution. We see that such a prior complements well the low distance map by adding spatially averaged large structures farther away. This means that data from statistical methods applied to massive photometric surveys and also future Gaia data may be used in conjunction with the databases of individual sightline data.}
\end{itemize}

The resulting 3D maps are significantly improved with respect to the previous maps thanks to the use of Gaia parallaxes, the additional DIB database, and the use of a prior. 
This is demonstrated by the residuals, and  especially by the increased compactness of the structures and the disappearance or decrease of some of the elongated artifacts due to contradictory distances (the \textit{fingers of god}) or lack of targets. 
Many of these artifacts remain, however, and are expected to disappear when all targets {are assigned} an accurate distance. 
In addition, the comparisons with other cloud location assignments and X-ray emitting cavities are encouraging. 
This shows that future Gaia parallax catalogs should considerably help the 3D mapping by allowing the use of a much larger number of targets.  
Merging photometric and DIB-based color excess estimates does not reveal any contradictory information and instead improves the maps, especially beyond foreground dense clouds. 
This opens the perspective of using the massive amounts of DIB data that are expected to be provided from high-resolution spectroscopic surveys. 
Finally, the use of a Pan-STARRS based prior demonstrates the feasibility of hierarchical methods for the inversion based on different datasets or massive amounts of data. 

Our distributions are particularly useful at short distances, as they show  how cloud complexes and cavities are spatially organized. 
In this respect, they are also appropriate tools for comparing with multiwavelength emission data, understanding environments of specific objects, computing particle or photon propagation modeled distributions, {and} searching for unreddened areas in case of stellar photometric model calibrations. 
In {terms} of angular resolution, they cannot compete with the distributions based on massive surveys. 
Especially, owing to the imposed cloud minimum size, opacities, and integrated reddenings toward the cores of the dense clouds can be strongly underestimated. 
In such nearby, opaque regions a valid approach may be the combination of {both types of information}, from high angular resolution maps, on the one hand, and from our 3D structure, on the other.

Vertical planar cuts in the 3D distribution are given in Appendix at longitude intervals of  30\fdeg. Our final map is available at stilism.obspm.fr in its entirety and online tools {allow one} to obtain reddening curves estimated through integration in the 3D opacity distribution along with rough estimates of the uncertainties on distances of the structures and reddening. We plan to update the maps and improve online tools when additional data are included in the inversion.

%We give some examples of these improvements in several planes. 
%Final map has a $\chi^2$ better then \citep{lallement14}, so we could affirm that we could create a map using photometric
% extinction and interstellar
%absorption at the same time with coherent results. Using  \prr{} we add global information about structures, and it helps to define 
%farer structures where we have less data. Our maps could be use with the \citep{Green15}, because our precision in the closer part and their one in the farer then 500 pc.
%Using DIBs equivalent width to trace the excess of colour is not precise as single value, but a good amount of data 
%increase the precision with good quality information. 
%Our 3D maps are available on the website ****.
%In the future, we hope to improve the maps with others absorption catalog as input, with the Hierarchical technique to have more 
%reliable structiures and probably an more precise idea about errors. The map should not used without the general contest of precision and 
%data distribution.
% 

\begin{acknowledgements}
We deeply thank our referee Katia Ferri{\'e}re for careful reading of the manuscript and numerous constructive comments. Her corrections and suggestions resulted in significant improvement of the article.
L.C. acknowledges doctoral grant funding from the Centre National d'Etudes Spatiales (CNES). 
R.L. and A.M.-I. acknowledge support from "Agence Nationale de la Recherche" through the STILISM project (ANR-12-BS05-0016-02) and the CNRS PCMI national program.
M.E. acknowledges funding from the "Region Ile-de-France" through the DIM-ACAV project.

This research has made use of the SIMBAD database, operated at CDS, Strasbourg, France.
\end{acknowledgements}

\bibliographystyle{aa} 
\bibliography{mybib}

\begin{thebibliography}{37}
\expandafter\ifx\csname natexlab\endcsname\relax\def\natexlab#1{#1}\fi

\bibitem[{{Aihara} {et~al.}(2011){Aihara}, {Allende Prieto}, {An}, {Anderson},
  {Aubourg}, {Balbinot}, {Beers}, {Berlind}, {Bickerton}, {Bizyaev}, {Blanton},
  {Bochanski}, {Bolton}, {Bovy}, {Brandt}, {Brinkmann}, {Brown}, {Brownstein},
  {Busca}, {Campbell}, {Carr}, {Chen}, {Chiappini}, {Comparat}, {Connolly},
  {Cortes}, {Croft}, {Cuesta}, {da Costa}, {Davenport}, {Dawson}, {Dhital},
  {Ealet}, {Ebelke}, {Edmondson}, {Eisenstein}, {Escoffier}, {Esposito},
  {Evans}, {Fan}, {Femen{\'{\i}}a Castell{\'a}}, {Font-Ribera}, {Frinchaboy},
  {Ge}, {Gillespie}, {Gilmore}, {Gonz{\'a}lez Hern{\'a}ndez}, {Gott}, {Gould},
  {Grebel}, {Gunn}, {Hamilton}, {Harding}, {Harris}, {Hawley}, {Hearty}, {Ho},
  {Hogg}, {Holtzman}, {Honscheid}, {Inada}, {Ivans}, {Jiang}, {Johnson},
  {Jordan}, {Jordan}, {Kazin}, {Kirkby}, {Klaene}, {Knapp}, {Kneib},
  {Kochanek}, {Koesterke}, {Kollmeier}, {Kron}, {Lampeitl}, {Lang}, {Le Goff},
  {Lee}, {Lin}, {Long}, {Loomis}, {Lucatello}, {Lundgren}, {Lupton}, {Ma},
  {MacDonald}, {Mahadevan}, {Maia}, {Makler}, {Malanushenko}, {Malanushenko},
  {Mandelbaum}, {Maraston}, {Margala}, {Masters}, {McBride}, {McGehee},
  {McGreer}, {M{\'e}nard}, {Miralda-Escud{\'e}}, {Morrison}, {Mullally},
  {Muna}, {Munn}, {Murayama}, {Myers}, {Naugle}, {Neto}, {Nguyen}, {Nichol},
  {O'Connell}, {Ogando}, {Olmstead}, {Oravetz}, {Padmanabhan},
  {Palanque-Delabrouille}, {Pan}, {Pandey}, {P{\^a}ris}, {Percival},
  {Petitjean}, {Pfaffenberger}, {Pforr}, {Phleps}, {Pichon}, {Pieri}, {Prada},
  {Price-Whelan}, {Raddick}, {Ramos}, {Reyl{\'e}}, {Rich}, {Richards}, {Rix},
  {Robin}, {Rocha-Pinto}, {Rockosi}, {Roe}, {Rollinde}, {Ross}, {Ross},
  {Rossetto}, {S{\'a}nchez}, {Sayres}, {Schlegel}, {Schlesinger}, {Schmidt},
  {Schneider}, {Sheldon}, {Shu}, {Simmerer}, {Simmons}, {Sivarani}, {Snedden},
  {Sobeck}, {Steinmetz}, {Strauss}, {Szalay}, {Tanaka}, {Thakar}, {Thomas},
  {Tinker}, {Tofflemire}, {Tojeiro}, {Tremonti}, {Vandenberg}, {Vargas
  Maga{\~n}a}, {Verde}, {Vogt}, {Wake}, {Wang}, {Weaver}, {Weinberg}, {White},
  {White}, {Yanny}, {Yasuda}, {Yeche}, \& {Zehavi}}]{Aihara11}
{Aihara}, H., {Allende Prieto}, C., {An}, D., {et~al.} 2011, ApJS, 193, 29

\bibitem[{{Alam} {et~al.}(2015){Alam}, {Albareti}, {Allende Prieto}, {Anders},
  {Anderson}, {Anderton}, {Andrews}, {Armengaud}, {Aubourg}, {Bailey}, \&
  et~al.}]{Alam15}
{Alam}, S., {Albareti}, F.~D., {Allende Prieto}, C., {et~al.} 2015, ApJS, 219,
  12

\bibitem[{{Arenou} {et~al.}(2017){Arenou}, {Luri}, {Babusiaux}, {Fabricius},
  {Helmi}, {Robin}, {Vallenari}, {Blanco-Cuaresma}, {Cantat-Gaudin},
  {Findeisen}, {Reyl{\'e}}, {Ruiz-Dern}, {Sordo}, {Turon}, {Walton}, {Shih},
  {Antiche}, {Barache}, {Barros}, {Breddels}, {Carrasco}, {Costigan},
  {Diakit{\'e}}, {Eyer}, {Figueras}, {Galluccio}, {Heu}, {Jordi},
  {Krone-Martins}, {Lallement}, {Lambert}, {Leclerc}, {Marrese}, {Moitinho},
  {Mor}, {Romero-G{\'o}mez}, {Sartoretti}, {Soria}, {Soubiran}, {Souchay},
  {Veljanoski}, {Ziaeepour}, {Giuffrida}, {Pancino}, \& {Bragaglia}}]{Arenou17}
{Arenou}, F., {Luri}, X., {Babusiaux}, C., {et~al.} 2017, \aap, 599, A50

\bibitem[{{Breitschwerdt} {et~al.}(2016){Breitschwerdt}, {Feige}, {Schulreich},
  {Avillez}, {Dettbarn}, \& {Fuchs}}]{Breitschwerdt16}
{Breitschwerdt}, D., {Feige}, J., {Schulreich}, M.~M., {et~al.} 2016, \nat,
  532, 73

\bibitem[{{Cramer}(1999)}]{Cramer99}
{Cramer}, N. 1999, \nar, 43, 343

\bibitem[{{Dias} {et~al.}(2012){Dias}, {Alessi}, {Moitinho}, \&
  {Lepine}}]{Dias12}
{Dias}, W.~S., {Alessi}, B.~S., {Moitinho}, A., \& {Lepine}, J.~R.~D. 2012,
  VizieR Online Data Catalog, 1

\bibitem[{{Eisenstein} {et~al.}(2011){Eisenstein}, {Weinberg}, {Agol},
  {Aihara}, {Allende Prieto}, {Anderson}, {Arns}, {Aubourg}, {Bailey},
  {Balbinot}, \& et~al.}]{Eisenstein11}
{Eisenstein}, D.~J., {Weinberg}, D.~H., {Agol}, E., {et~al.} 2011, AJ, 142, 72

\bibitem[{{Elyajouri} {et~al.}(2017){Elyajouri}, {Lallement}, {Monreal-Ibero},
  {Capitanio}, \& {Cox}}]{Elyajouri17}
{Elyajouri}, M., {Lallement}, R., {Monreal-Ibero}, A., {Capitanio}, L., \&
  {Cox}, N.~L.~J. 2017, \aap, 600, A129

\bibitem[{{Elyajouri} {et~al.}(2016){Elyajouri}, {Monreal-Ibero}, {Remy}, \&
  {Lallement}}]{Elyajouri16}
{Elyajouri}, M., {Monreal-Ibero}, A., {Remy}, Q., \& {Lallement}, R. 2016,
  \apjs, 225, 19

\bibitem[{{Green} {et~al.}(2015){Green}, {Schlafly}, {Finkbeiner}, {Rix},
  {Martin}, {Burgett}, {Draper}, {Flewelling}, {Hodapp}, {Kaiser}, {Kudritzki},
  {Magnier}, {Metcalfe}, {Price}, {Tonry}, \& {Wainscoat}}]{Green15}
{Green}, G.~M., {Schlafly}, E.~F., {Finkbeiner}, D.~P., {et~al.} 2015, \apj,
  810, 25

\bibitem[{{Kos} {et~al.}(2014){Kos}, {Zwitter}, {Wyse}, {Bienaym{\'e}},
  {Binney}, {Bland-Hawthorn}, {Freeman}, {Gibson}, {Gilmore}, {Grebel},
  {Helmi}, {Kordopatis}, {Munari}, {Navarro}, {Parker}, {Reid}, {Seabroke},
  {Sharma}, {Siebert}, {Siviero}, {Steinmetz}, {Watson}, \& {Williams}}]{kos14}
{Kos}, J., {Zwitter}, T., {Wyse}, R., {et~al.} 2014, Science, 345, 791

\bibitem[{{Lallement}(2015)}]{Lall47tuc}
{Lallement}, R. 2015, in Journal of Physics Conference Series, Vol. 577,
  Journal of Physics Conference Series, 012016

\bibitem[{{Lallement} {et~al.}(2014){Lallement}, {Vergely}, {Valette},
  {Puspitarini}, {Eyer}, \& {Casagrande}}]{lallement14}
{Lallement}, R., {Vergely}, J.-L., {Valette}, B., {et~al.} 2014, A\&A, 561, A91

\bibitem[{{Lan} {et~al.}(2015){Lan}, {M{\'e}nard}, \& {Zhu}}]{Lan15}
{Lan}, T.-W., {M{\'e}nard}, B., \& {Zhu}, G. 2015, MNRAS, 452, 3629

\bibitem[{{Marshall} {et~al.}(2006){Marshall}, {Robin}, {Reyl{\'e}},
  {Schultheis}, \& {Picaud}}]{Marshall06}
{Marshall}, D.~J., {Robin}, A.~C., {Reyl{\'e}}, C., {Schultheis}, M., \&
  {Picaud}, S. 2006, \aap, 453, 635

\bibitem[{{Monreal-Ibero} \& {Lallement}(2015)}]{EWASS15}
{Monreal-Ibero}, A. \& {Lallement}, R. 2015, in Memorie della Societa Italiana
  Astronomica, Vol.~86, Geeting ready for Gaia:3D structure of the ISM,
  515--645

\bibitem[{{Nordstr{\"o}m} {et~al.}(2004){Nordstr{\"o}m}, {Mayor}, {Andersen},
  {Holmberg}, {Pont}, {J{\o}rgensen}, {Olsen}, {Udry}, \&
  {Mowlavi}}]{Nordstrom04}
{Nordstr{\"o}m}, B., {Mayor}, M., {Andersen}, J., {et~al.} 2004, \aap, 418, 989

\bibitem[{{Perrot} \& {Grenier}(2003)}]{Perrot03}
{Perrot}, C.~A. \& {Grenier}, I.~A. 2003, \aap, 404, 519

\bibitem[{{Puspitarini} {et~al.}(2015){Puspitarini}, {Lallement}, {Babusiaux},
  {Chen}, {Bonifacio}, {Sbordone}, {Caffau}, {Duffau}, {Hill}, {Monreal-Ibero},
  {Royer}, {Arenou}, {Peralta}, {Drew}, {Bonito}, {Lopez-Santiago}, {Alfaro},
  {Bensby}, {Bragaglia}, {Flaccomio}, {Lanzafame}, {Pancino}, {Recio-Blanco},
  {Smiljanic}, {Costado}, {Lardo}, {de Laverny}, \& {Zwitter}}]{Puspitarini15}
{Puspitarini}, L., {Lallement}, R., {Babusiaux}, C., {et~al.} 2015, A\&A, 573,
  A35

\bibitem[{{Puspitarini} {et~al.}(2014){Puspitarini}, {Lallement}, {Vergely}, \&
  {Snowden}}]{Puspitarini14}
{Puspitarini}, L., {Lallement}, R., {Vergely}, J.-L., \& {Snowden}, S.~L. 2014,
  \aap, 566, A13

\bibitem[{{Rezaei Kh.} {et~al.}(2016){Rezaei Kh.}, {Bailer-Jones}, {Hanson}, \&
  {Fouesneau}}]{Rezaei16}
{Rezaei Kh.}, S., {Bailer-Jones}, C.~A.~L., {Hanson}, R.~J., \& {Fouesneau}, M.
  2016, ArXiv e-prints [\eprint[arXiv]{1609.08917}]

\bibitem[{{Sale} \& {Magorrian}(2014{\natexlab{a}})}]{Sale14}
{Sale}, S.~E. \& {Magorrian}, J. 2014{\natexlab{a}}, MNRAS, 445, 256

\bibitem[{{Sale} \& {Magorrian}(2014{\natexlab{b}})}]{SaleMag14}
{Sale}, S.~E. \& {Magorrian}, J. 2014{\natexlab{b}}, MNRAS, 445, 256

\bibitem[{{Savage} \& {Mathis}(1979)}]{Savage79}
{Savage}, B.~D. \& {Mathis}, J.~S. 1979, \araa, 17, 73

\bibitem[{{Schlafly} {et~al.}(2014){Schlafly}, {Green}, {Finkbeiner}, {Rix},
  {Bell}, {Burgett}, {Chambers}, {Draper}, {Hodapp}, {Kaiser}, {Magnier},
  {Martin}, {Metcalfe}, {Price}, \& {Tonry}}]{Schlafly14}
{Schlafly}, E.~F., {Green}, G., {Finkbeiner}, D.~P., {et~al.} 2014, \apj, 786,
  29

\bibitem[{{Schlafly} {et~al.}(2015){Schlafly}, {Green}, {Finkbeiner}, {Rix},
  {Burgett}, {Chambers}, {Draper}, {Kaiser}, {Martin}, {Metcalfe}, {Morgan},
  {Price}, {Tonry}, {Wainscoat}, \& {Waters}}]{Schlafly15}
{Schlafly}, E.~F., {Green}, G., {Finkbeiner}, D.~P., {et~al.} 2015, \apj, 799,
  116

\bibitem[{{Schlegel} {et~al.}(1998){Schlegel}, {Finkbeiner}, \&
  {Davis}}]{Schlegel98}
{Schlegel}, D.~J., {Finkbeiner}, D.~P., \& {Davis}, M. 1998, \apj, 500, 525

\bibitem[{{Schultheis} {et~al.}(2014){Schultheis}, {Chen}, {Jiang}, {Gonzalez},
  {Enokiya}, {Fukui}, {Torii}, {Rejkuba}, \& {Minniti}}]{Schultheis14}
{Schultheis}, M., {Chen}, B.~Q., {Jiang}, B.~W., {et~al.} 2014, \aap, 566, A120

\bibitem[{{Skrutskie} {et~al.}(2006){Skrutskie}, {Cutri}, {Stiening},
  {Weinberg}, {Schneider}, {Carpenter}, {Beichman}, {Capps}, {Chester},
  {Elias}, {Huchra}, {Liebert}, {Lonsdale}, {Monet}, {Price}, {Seitzer},
  {Jarrett}, {Kirkpatrick}, {Gizis}, {Howard}, {Evans}, {Fowler}, {Fullmer},
  {Hurt}, {Light}, {Kopan}, {Marsh}, {McCallon}, {Tam}, {Van Dyk}, \&
  {Wheelock}}]{Skrutskie06}
{Skrutskie}, M.~F., {Cutri}, R.~M., {Stiening}, R., {et~al.} 2006, AJ, 131,
  1163

\bibitem[{{Tarantola} \& {Valette}(1982)}]{Tarantola82}
{Tarantola}, A. \& {Valette}, B. 1982, Reviews of Geophysics and Space Physics,
  20, 219

\bibitem[{{van Loon} {et~al.}(2013){van Loon}, {Bailey}, {Tatton}, {Ma{\'{\i}}z
  Apell{\'a}niz}, {Crowther}, {de Koter}, {Evans}, {H{\'e}nault-Brunet},
  {Howarth}, {Richter}, {Sana}, {Sim{\'o}n-D{\'{\i}}az}, {Taylor}, \&
  {Walborn}}]{vanloon13}
{van Loon}, J.~T., {Bailey}, M., {Tatton}, B.~L., {et~al.} 2013, A\&A, 550,
  A108

\bibitem[{{Vergely} {et~al.}(2001){Vergely}, {Freire Ferrero}, {Siebert}, \&
  {Valette}}]{Vergely01}
{Vergely}, J.-L., {Freire Ferrero}, R., {Siebert}, A., \& {Valette}, B. 2001,
  \aap, 366, 1016

\bibitem[{{Vergely} {et~al.}(2010){Vergely}, {Valette}, {Lallement}, \&
  {Raimond}}]{Vergely10}
{Vergely}, J.-L., {Valette}, B., {Lallement}, R., \& {Raimond}, S. 2010, A\&A,
  518, A31

\bibitem[{{Welsh} {et~al.}(2010){Welsh}, {Lallement}, {Vergely}, \&
  {Raimond}}]{Welsh10}
{Welsh}, B.~Y., {Lallement}, R., {Vergely}, J.-L., \& {Raimond}, S. 2010, \aap,
  510, A54

\bibitem[{{Wilson} {et~al.}(2010){Wilson}, {Hearty}, {Skrutskie}, {Majewski},
  {Schiavon}, {Eisenstein}, {Gunn}, {Blank}, {Henderson}, {Smee}, {Barkhouser},
  {Harding}, {Fitzgerald}, {Stolberg}, {Arns}, {Nelson}, {Brunner}, {Burton},
  {Walker}, {Lam}, {Maseman}, {Barr}, {Leger}, {Carey}, {MacDonald}, {Horne},
  {Young}, {Rieke}, {Rieke}, {O'Brien}, {Hope}, {Krakula}, {Crane}, {Zhao},
  {Carr}, {Harrison}, {Stoll}, {Vernieri}, {Holtzman}, {Shetrone},
  {Allende-Prieto}, {Johnson}, {Frinchaboy}, {Zasowski}, {Bizyaev},
  {Gillespie}, \& {Weinberg}}]{Wilson10}
{Wilson}, J.~C., {Hearty}, F., {Skrutskie}, M.~F., {et~al.} 2010, in Society of
  Photo-Optical Instrumentation Engineers (SPIE) Conference Series, Vol. 7735,
  Society of Photo-Optical Instrumentation Engineers (SPIE) Conference Series,
  1

\bibitem[{{Zasowski} {et~al.}(2013){Zasowski}, {Johnson}, {Frinchaboy},
  {Majewski}, {Nidever}, {Rocha Pinto}, {Girardi}, {Andrews}, {Chojnowski},
  {Cudworth}, {Jackson}, {Munn}, {Skrutskie}, {Beaton}, {Blake}, {Covey},
  {Deshpande}, {Epstein}, {Fabbian}, {Fleming}, {Garcia Hernandez}, {Herrero},
  {Mahadevan}, {M{\'e}sz{\'a}ros}, {Schultheis}, {Sellgren}, {Terrien}, {van
  Saders}, {Allende Prieto}, {Bizyaev}, {Burton}, {Cunha}, {da Costa},
  {Hasselquist}, {Hearty}, {Holtzman}, {Garc{\'{\i}}a P{\'e}rez}, {Maia},
  {O'Connell}, {O'Donnell}, {Pinsonneault}, {Santiago}, {Schiavon}, {Shetrone},
  {Smith}, \& {Wilson}}]{Zasowski13}
{Zasowski}, G., {Johnson}, J.~A., {Frinchaboy}, P.~M., {et~al.} 2013, AJ, 146,
  81

\bibitem[{{Zasowski} {et~al.}(2015){Zasowski}, {M{\'e}nard}, {Bizyaev},
  {Garc{\'{\i}}a-Hern{\'a}ndez}, {Garc{\'{\i}}a P{\'e}rez}, {Hayden},
  {Holtzman}, {Johnson}, {Kinemuchi}, {Majewski}, {Nidever}, {Shetrone}, \&
  {Wilson}}]{zasowski15}
{Zasowski}, G., {M{\'e}nard}, B., {Bizyaev}, D., {et~al.} 2015, ApJ, 798, 35

\end{thebibliography}

\section{Appendix}

\begin{figure*}
\centering
\includegraphics[width=.5\textwidth]{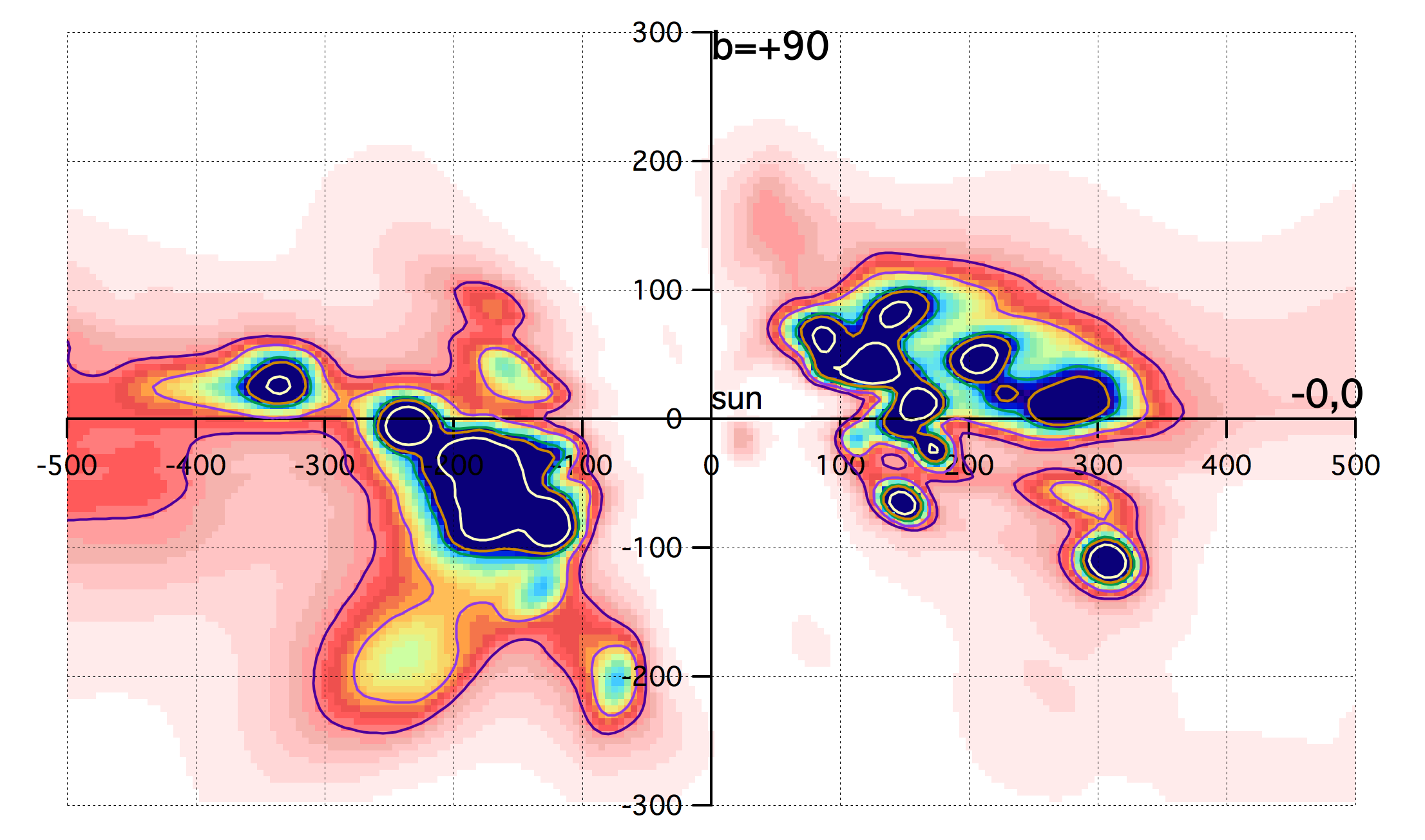}
\caption{Vertical cut in the 3D opacity distribution  {for D inversion} along the meridian plane l=0-180\fdeg. Color scale and iso-contours as in Fig \ref{cut212}.}
\label{vercut0}
\end{figure*}

\begin{figure*}
\centering
\includegraphics[width=.5\textwidth]{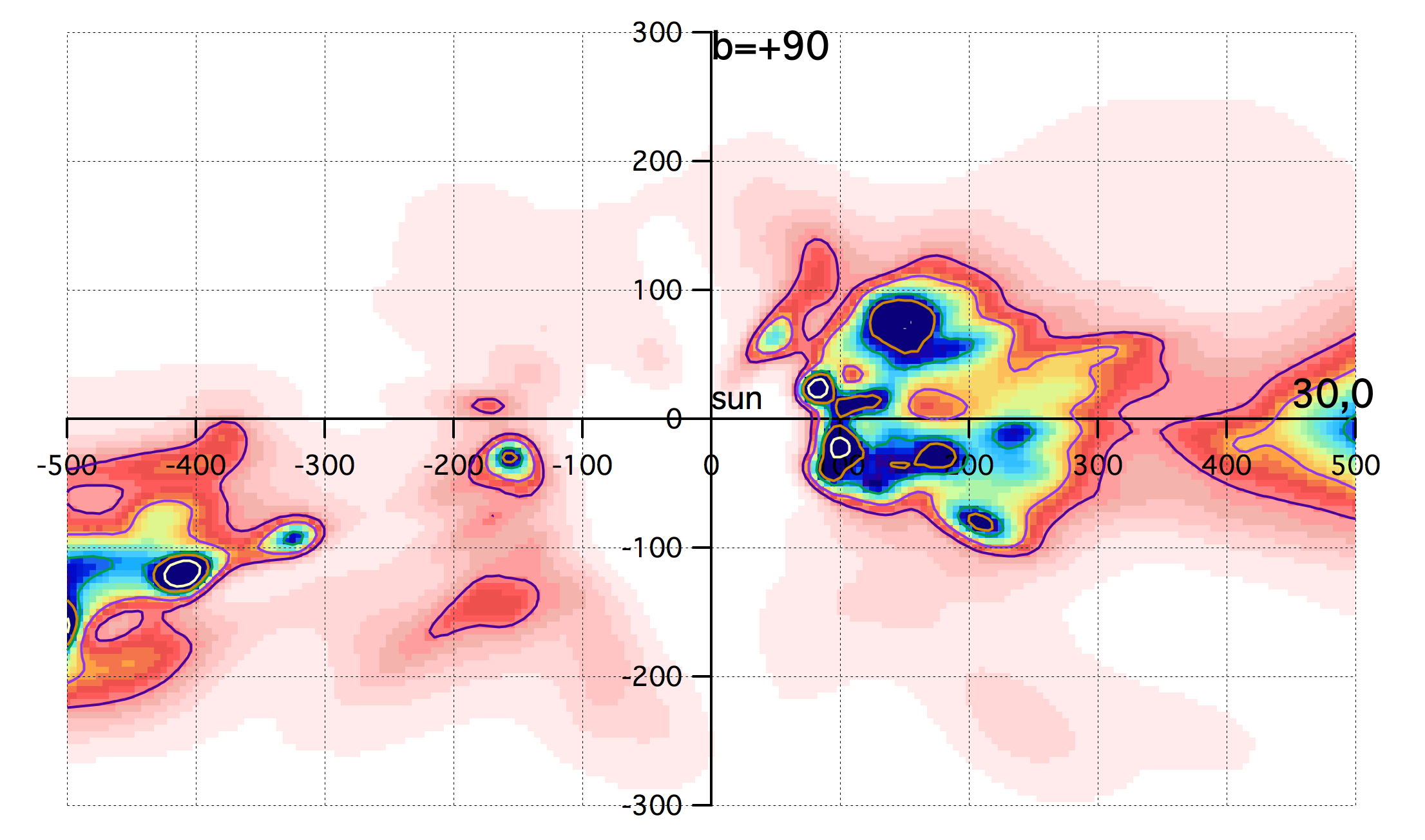}
\caption{Same as Fig. \ref{vercut0} for the vertical plane along l=30-210\fdeg}
\label{vercut30}
\end{figure*}

\begin{figure*}
\centering
\includegraphics[width=.5\textwidth]{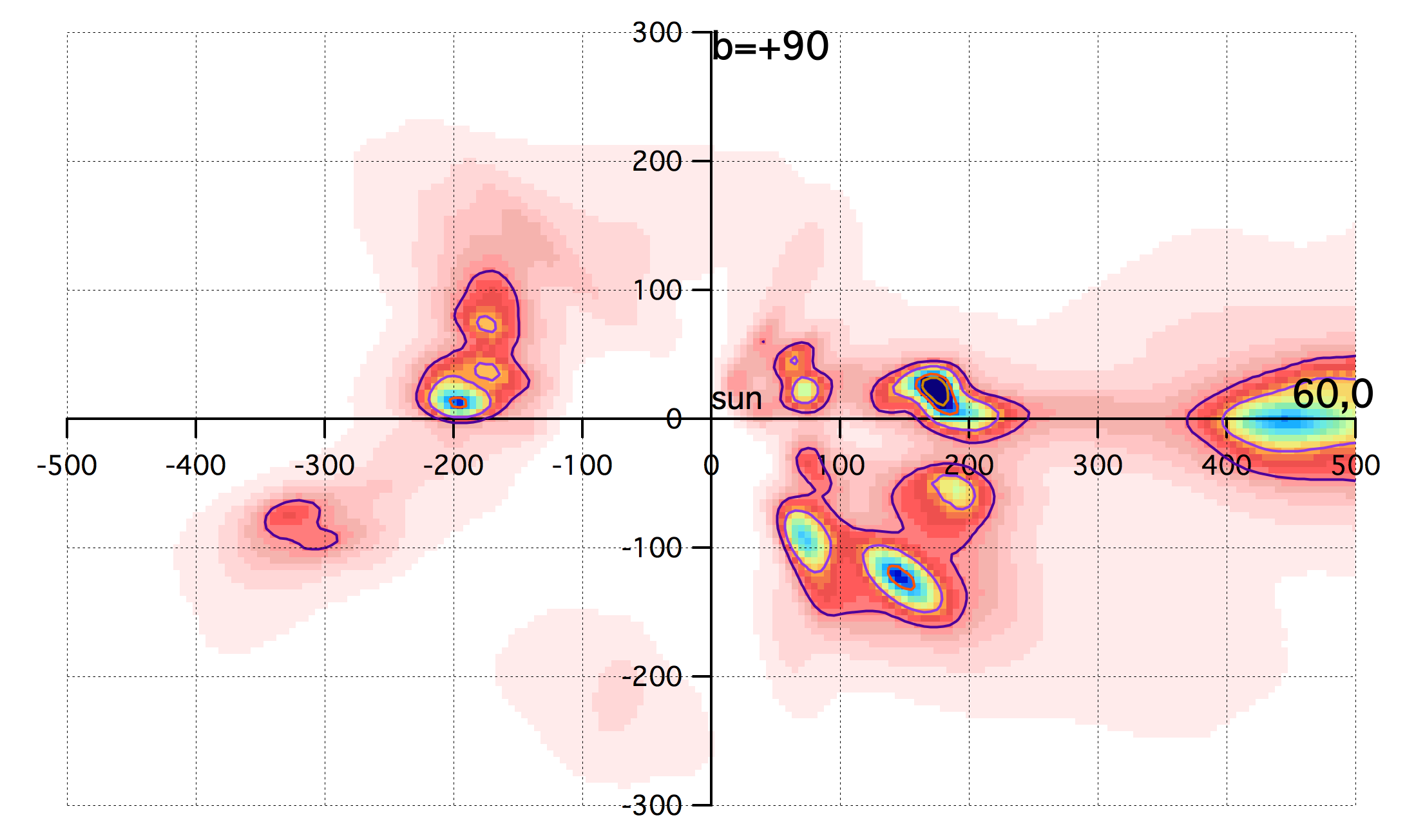}
\caption{Same as Fig. \ref{vercut0} for the vertical plane l=60-240\fdeg}
\label{vercut60}
\end{figure*}

\begin{figure*}
\centering
\includegraphics[width=.5\textwidth]{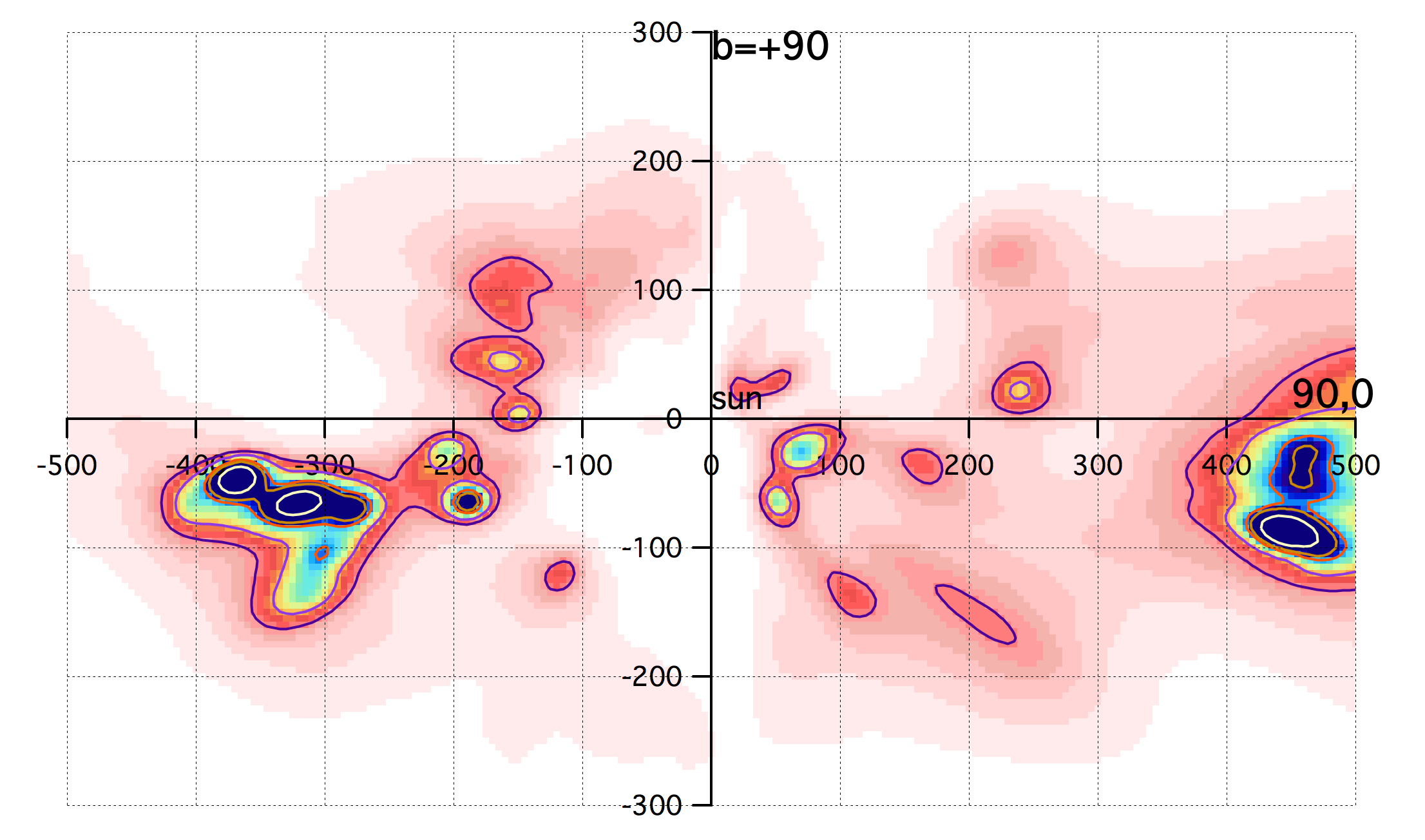}
\caption{Same as Fig. \ref{vercut0} for the vertical plane l=90-270\fdeg (the rotation plane)}
\label{vercut90}
\end{figure*}

\begin{figure*}
\centering
\includegraphics[width=.5\textwidth]{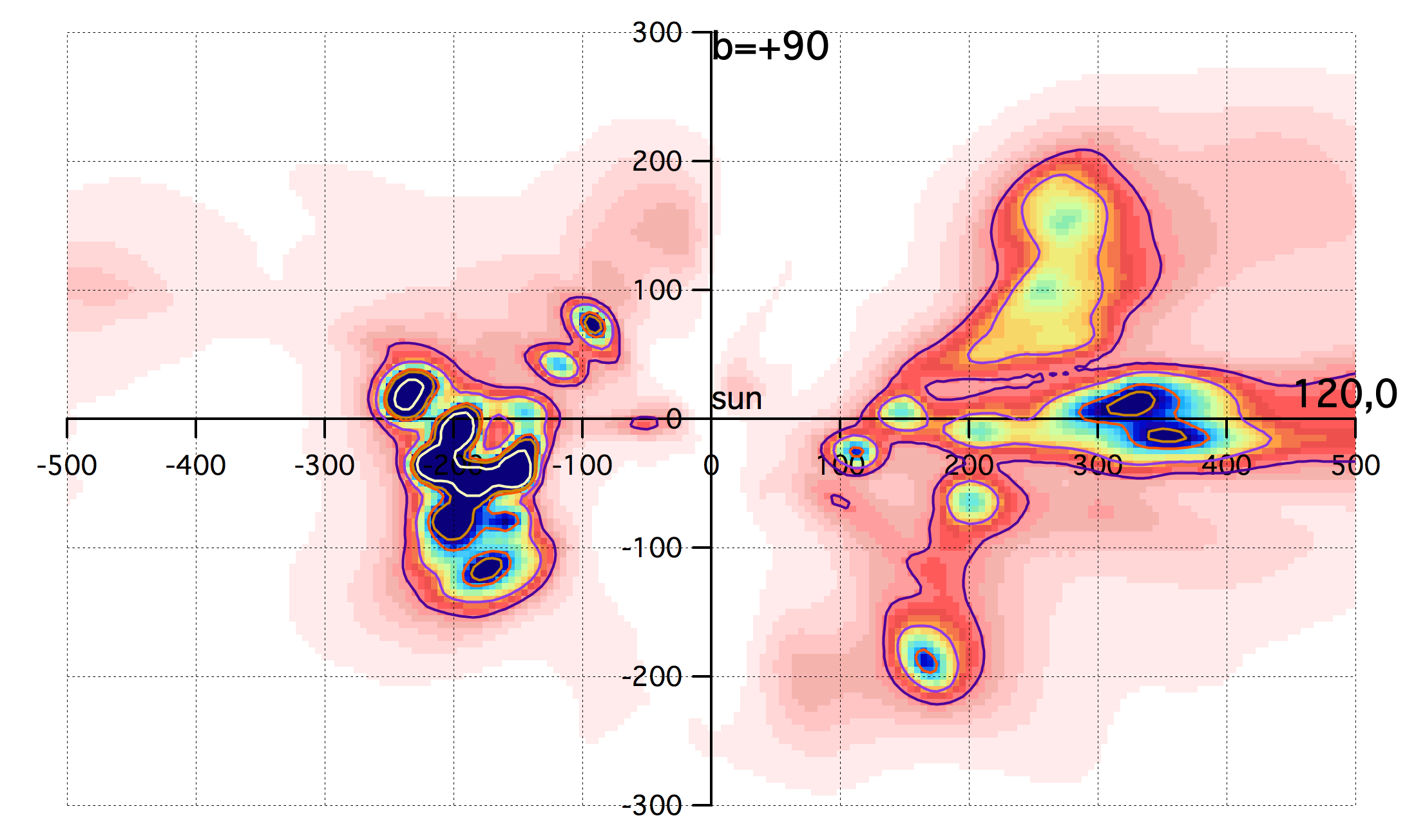}
\caption{Same as Fig. \ref{vercut0} for the vertical plane l=120-300$\fdeg$.}
\label{vercut120}
\end{figure*}

\begin{figure*}
\centering
\includegraphics[width=.5\textwidth]{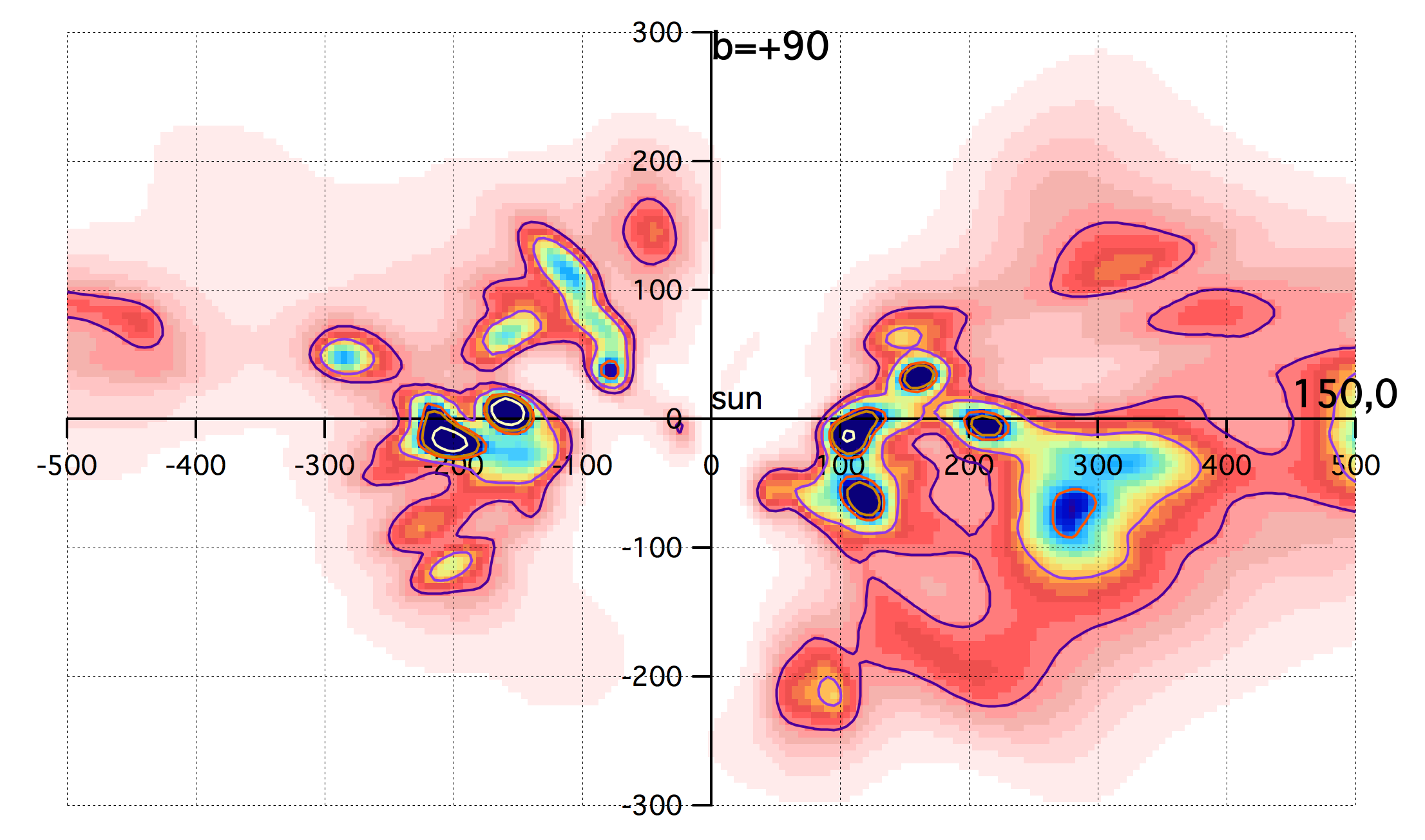}
\caption{Same as Fig. \ref{vercut0} for the vertical plane l=150-330\fdeg.}
\label{vercut150}
\end{figure*}
\end{document}